\documentclass{aastex631}
\usepackage{natbib}
\usepackage{xcolor}

\usepackage{amsmath}
\usepackage{wrapfig}
\newcommand \kms{km~$\rm{s}^{-1}$}

\graphicspath{{./}{figures/}}

\begin{document}
\title{Interstellar planetesimals}

\author{Amaya Moro-Mart\'{\i}n}
\affiliation{Space Telescope Science Institute, 3700 San Mart{\'\i}n Dr., Baltimore, MD 21218}
\affiliation{Physics and Astronomy, Center for Astrophysical Sciences, Johns Hopkins University, Baltimore, MD 21218}

\begin{abstract}
During the formation of our solar system, a large number of planetesimals were ejected into interstellar space by gravitational encounters with the planets. Debris disks observations and numerical simulations indicate that many other planetary systems, now known to be quite common, would have undergone a similar dynamical clearing process. It is therefore expected that the galaxy should be teeming with expelled planetesimals, largely unaltered since their ejection. This is why astronomers were perplexed that none had been detected passing through the solar system. Then, in 2017, the discovery of1I/'Oumuamua transformed the situation from puzzlement to bewilderment. Its brief visit and limited observations left important questions about its nature and origin unanswered and raised the possibility that 1I/'Oumuamua could be a never-seen-before intermediate product of planet formation. If so, this could open a new observational window to study the primordial building blocks of planets, setting unprecedented constraints on planet formation models. Two years later 2I/Borisov was discovered, with an unquestionable cometary composition, confirming that a population of icy interstellar planetesimals exists. These objects have remained largely unchanged since their ejection, like time capsules of their planetary system most distant past. Interstellar planetesimals could potentially be trapped into star and planet formation environments, acting as seeds for planet formation, helping overcome the meter-size barrier that challenges the growth of cm-sized pebbles into km-sized objects. Interstellar planetesimals play a pivotal role in our understanding of planetary system formation and evolution and point to the possibility that one day, we will be able to hold a fragment from another world in our hand.
\end{abstract}

\section{An astounding yet expected discovery}
\label{discovery}

It was 18 October 2017 when the {\it Panoramic Survey Telescope and Rapid Response System} ({\it Pan-STARRS}), a 1.8 m telescope that it is constantly surveying the sky for objects that are either moving or that are variable, made an astounding discovery. That day, University of Hawaii astronomer Robert Weryk identified an object (P10Ee5V) with a very unusual orbit that was later confirmed to be hyperbolic (Williams et al. 2017). Figure 1 shows its trajectory as it entered the inner solar system. Like a bat appearing from nowhere, it had sneaked in from above the ecliptic plane, sank below the plane, and on 14 October 2017 re-emerged remarkably close to the Earth, just 60 times the distance to the Moon, revealing itself for the first time shortly after. The astronomical community was immediately notified and many observatories rushed to observe it. Unfortunately, it was already on its way out of the solar system, having passed closest to the Sun on 9 September 2017 at a distance of 0.25 au.  It was last seen with {\it HST} on 2 January 2019, leaving astronomers astounded and yearning for more data. Studies based on visible/near-infrared observations lead to an effective radius\footnote{Radius estimates are 55 m (for albedo 0.1, Jewitt et al. \citeyear{2017ApJ...850L..36J}), 60 m (for albedo 0.04, Fraser et al. \citeyear{2018NatAs...2..383F}), 80 m (for albedo of 0.037, Drahus et al. \citeyear{2018NatAs...2..407D}), 102 m (for albedo 0.04, Meech et al. \citeyear{2017Natur.552..378M}), and 130 m (for albedo 0.03, Bolin et al. \citeyear{2018ApJ...852L...2B}).} in the range of 55--130 m (Jewitt et al. \citeyear{2017ApJ...850L..36J}; Meech et al. \citeyear{2017Natur.552..378M}; Fraser et al. \citeyear{2018NatAs...2..383F}; Drahus et al. \citeyear{2018NatAs...2..407D}; Bolin et al. \citeyear{2018ApJ...852L...2B}).  Observations with the {\it Spitzer Space Telescope} on 21-22 November 2017 could not detect its thermal emission, with the 3$\sigma$ upper limit at 4.5 $\mu$m leading to an effective spherical radius of less than [49, 70, 220] m and albedo greater than [0.2, 0.1, 0.01] (Trilling et al. \citeyear{2018AJ....156..261T}). This first interstellar interloper was a remarkably small object that was fortuitously detected in a magnitude-limited survey after its perihelion passage because it happened to pass very close to the Earth on its way out of the solar system (Jewitt et al. \citeyear{2017ApJ...850L..36J}). Only a few solar system objects of that size have been studied and they are all near-Earth objects, limiting our ability to make comparisons. 

Its brightness was found to fluctuate very sharply every 4 hours with a rotation period of about 8 hours. Some brightness variation is expected from irregularly shaped objects because as they rotate their cross section vary; it is also expected  if the objects have a non-uniform albedo, showing at times areas that may be more reflective than others. But for solar system objects this brightness variation is at a few percent level whereas for this first interstellar object it was about a factor of ten. This was interpreted as evidence of a morphology that was unusually elongated, with an axis ratio\footnote{1I/'Oumuamua axis ratio estimates are $>$ 6:1 in Jewitt et al. (\citeyear{2017ApJ...850L..36J}); 5:3.1 in Banninster et al. (\citeyear{2017ApJ...851L..38B}); 10:1 in Meech et al. (\citeyear{2017Natur.552..378M}); 3:1 in Knight et al. (\citeyear{2017ApJ...851L..31K}); $>$ 4.63 in Drahus et al. (\citeyear{2018NatAs...2..407D}); from 3.5 to 10.3 in Bolin et al. (\citeyear{2018ApJ...852L...2B}); $>$ 5:1 in Fraser et al. (\citeyear{2018NatAs...2..383F}); 6$\pm$1:1 in McNeill et al. (\citeyear{2018ApJ...857L...1M}).} ranging from 3 to 10 (Jewitt et al. \citeyear{2017ApJ...850L..36J};  Banninster et al. \citeyear{2017ApJ...851L..38B};  Meech et al. \citeyear{2017Natur.552..378M};  Knight et al. \citeyear{2017ApJ...851L..31K};  Drahus et al. \citeyear{2018NatAs...2..407D};  Bolin et al. \citeyear{2018ApJ...852L...2B};  Fraser et al. \citeyear{2018NatAs...2..383F};  McNeill et al. \citeyear{2018ApJ...857L...1M}), or oblate (Belton et al. \citeyear{2018ApJ...856L..21B}). It has been proposed that the elongated shape is more likely because is energetically more stable and requires less fine-tuned orientations to explain its lightcurve (Oumuamua ISSI Team et al. \citeyear{2019NatAs...3..594O}). Unfortunately, its shape will remain unknown because no period solution was agreed upon (with a possible non-principal axis rotation), there were not enough observations sampling different phase angles, and also because it is unknown to what degree surface albedo variations contributed to the observed flux variability. Figure \ref{1Imorphology} shows some of the proposed morphologies. 

Its red color, based on photometric and spectral data, is consistent with iron-rich minerals and with space weathered surfaces and resembled small bodies in the solar system, including D-type asteroids, some Jupiter Trojans and trans-Neptunian objects, and comets (Meech et al. \citeyear{2017Natur.552..378M}; Jewitt et al. \citeyear{2017ApJ...850L..36J}; Banninster et al. \citeyear{2017ApJ...851L..38B}; Bolin et al. \citeyear{2018ApJ...852L...2B}; Fitzsimmons et al. \citeyear{2018NatAs...2..133F}; Oumuamua ISSI Team et al. \citeyear{2019NatAs...3..594O}). 

The different denominations that this first interstellar object received reflect the initial confusion regarding its origin. Initially renamed C/2017 U1 (where C/ refers to comets that are not periodic), it was soon renamed A/2017 U1 (where A/ refers to asteroid) because, in spite its close approach to the Sun, the object did not show any dust or gas outflow characteristic of cometary activity (Jewitt et al. \citeyear{2017ApJ...850L..36J}; Meech et al. \citeyear{2017Natur.552..378M}; Ye et al. \citeyear{2017ApJ...851L...5Y}; Trilling et al. \citeyear{2018AJ....156..261T}). The A/  denomination did not last long either because its orbit was unquestionably hyperbolic, like none of the solar system objects. Its current name is 1I/2017 U1, where I/ refers to all interstellar objects, whether cometary or asteroidal in nature. It is also known as 1I/'Oumuamua, that comes from the Hawaiian $'ou$ (reach out for) and $mua$ (first, in advance of). 

Despite the surprise, the arrival of this first interstellar visitor was not unexpected. On 30 August 2019 a second interstellar interloper was detected by amateur astronomer and optical engineer Guennadi Borisov (Borisov 2019), using a homemade 0.65 m telescope. Figure \ref{2Itrajectory} shows 2I/Borisov's trajectory. 

The detections of 1I/'Oumuamua and 2I/Borisov have opened a new era in astronomy because never before have we been able to study "up close" objects from outside our solar system. In Section \ref{planetformation} we will see that, even more extraordinary, these objects most likely originate from extrasolar planetary systems and have remained largely unchanged since their ejection, like time capsules of their planetary system most distant past. 

\begin{figure}
\begin{center}
\includegraphics[width=10cm]{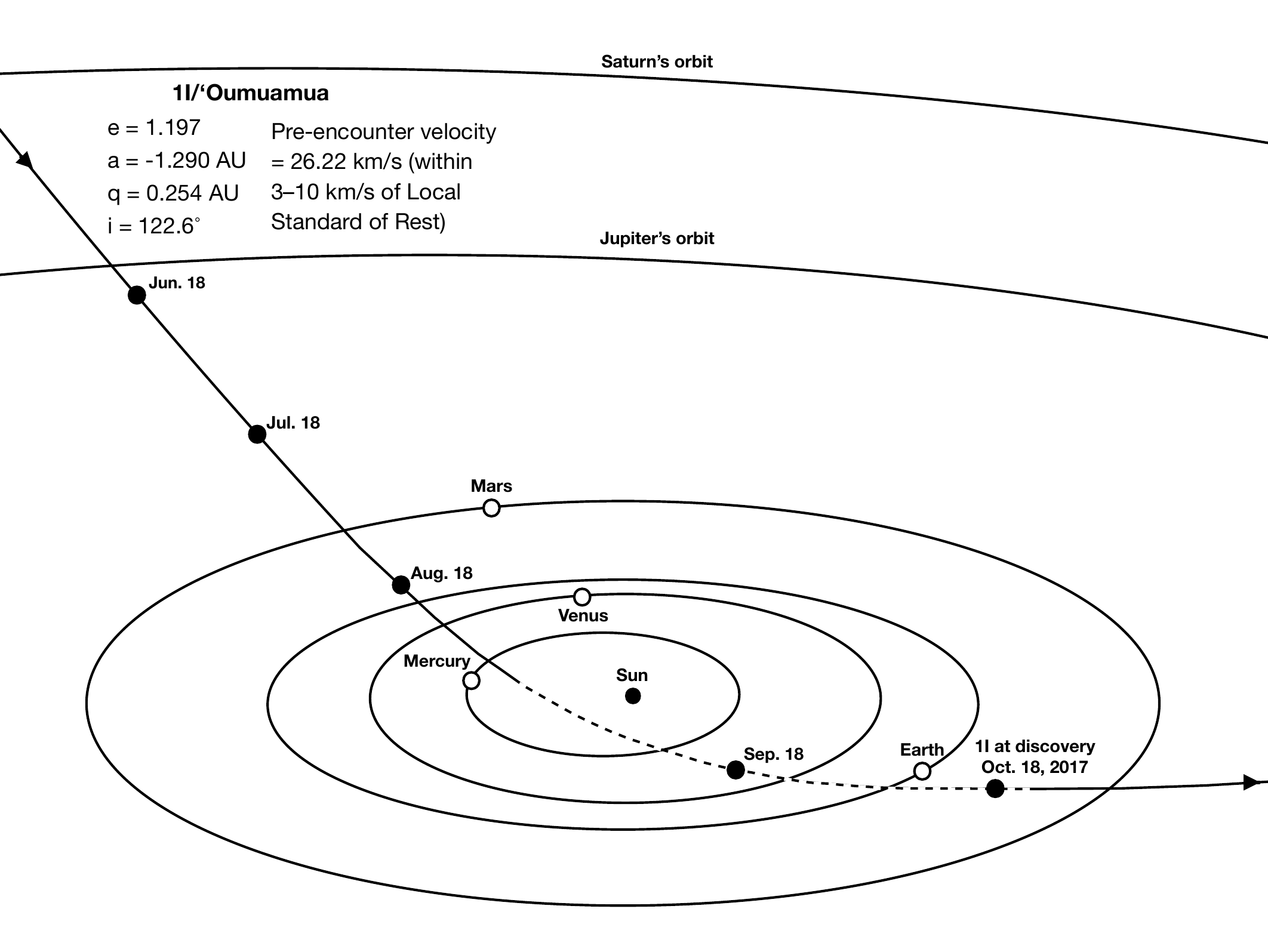}
\end{center}
\caption{1I/'Oumuamua's trajectory as it entered the inner solar system (dashed line indicates the section that lies below the ecliptic plane). The open circles show the position of the planets at the time of 1I/'Oumuamua's discovery. Based on a figure by Matthew Twombly for Jewitt \& Moro-Mart{\'\i}n (2020).}
\label{1Itrajectory}
\end{figure}

\begin{figure}
\begin{center}
\includegraphics[width=10cm]{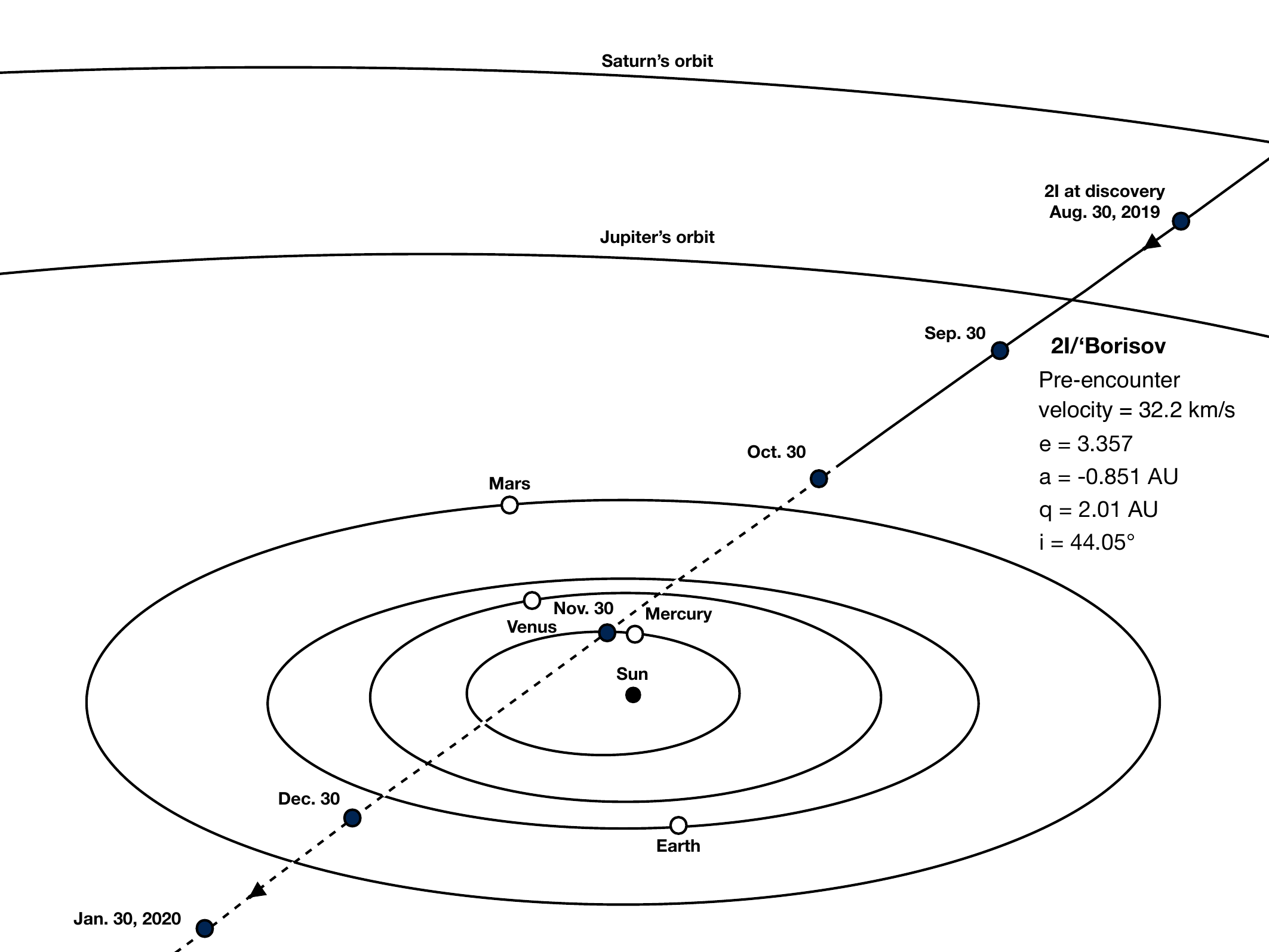}
\end{center}
\caption{2I/Borisov's trajectory as it entered the inner Solar system (dashed line indicates the section that lies below the ecliptic plane). The open circles show the position of the planets at the time of 2I/Borisov discovery. Based on a figure by Matthew Twombly for Jewitt \& Moro-Mart{\'\i}n (2020).}
\label{2Itrajectory}
\end{figure}

\begin{figure}
\begin{center}
\includegraphics[width=3cm]{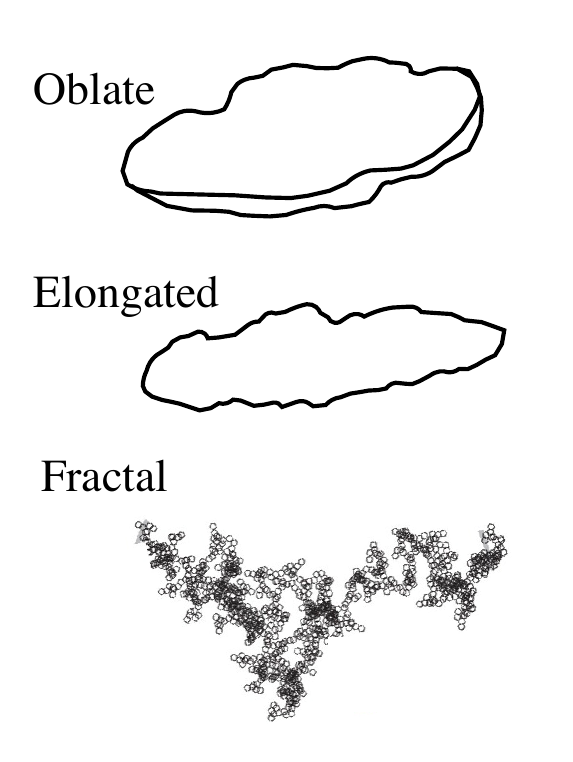}
\end{center}
\caption{Some of the morphologies that have been suggested for 1I/'Oumuamua.}
\label{1Imorphology}
\end{figure}

\section{Interstellar planetesimals are a byproduct of planet formation and evolution}
\label{planetformation}

\begin{figure}
\begin{center}
\includegraphics[width=9cm]{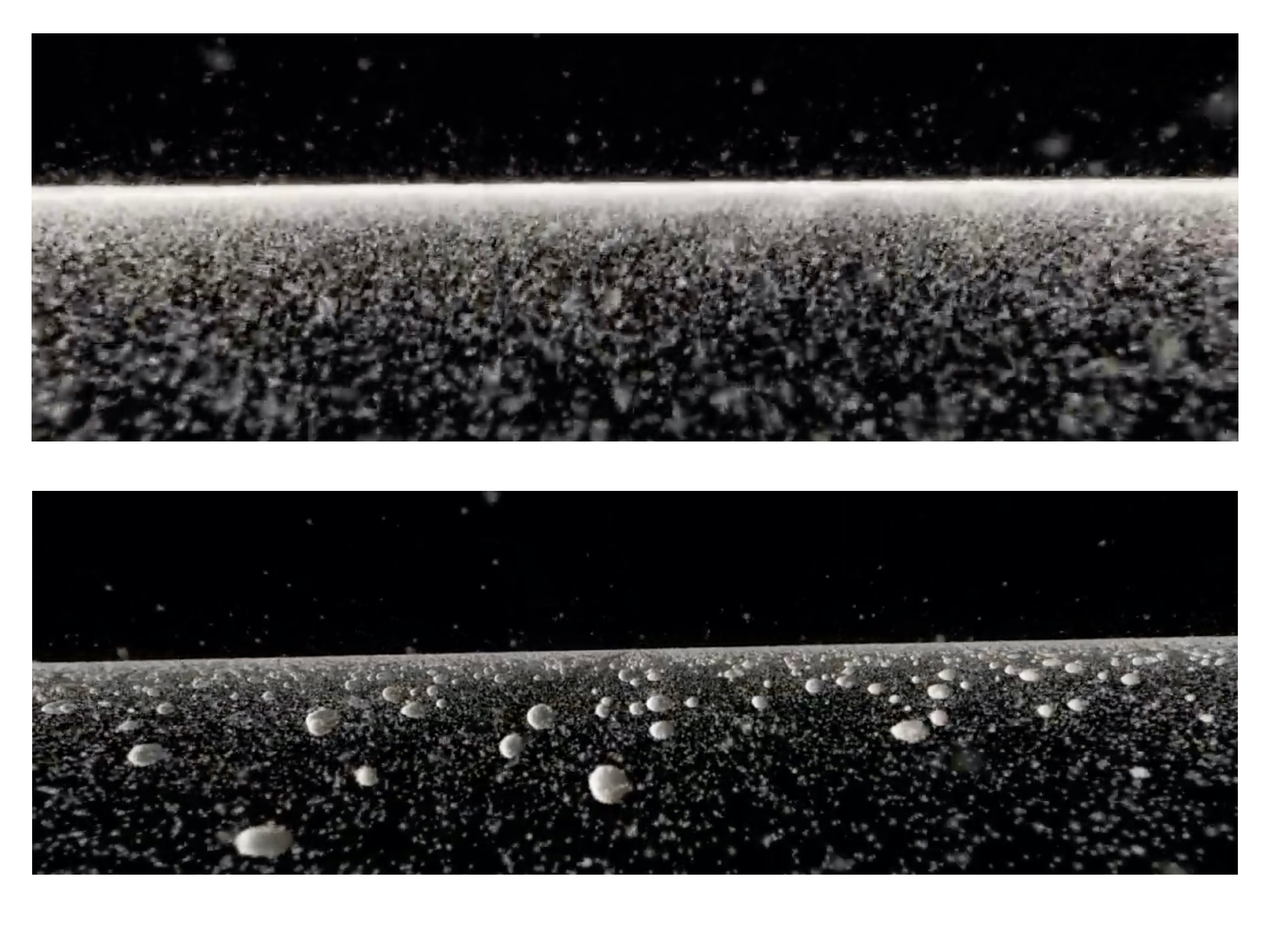}
\end{center}
\caption{Snapshots from a numerical simulation of planetesimal formation in a protoplanetary disk, viewed from above the plane. Credit: Shugo Michikoshi, Eiichiro Kokubo, Hirotaka Nakayama, Yayoi Narazaki, 4D2U Project, NAOJ.}
\label{Kokubo}
\end{figure}

Stars form from the collapse of dense regions of molecular clouds. A natural byproduct of this process is the formation of protoplanetary disks, where planet formation takes place (Hayashi \citeyear{1981PThPS..70...35H}; Shu et al. \citeyear{1987ARA&A..25...23S}; Hartmann \citeyear{Hartmann2008}). The dust particles in these disks, approximately 0.1 $\mu$m in size, are strongly coupled to the gas, resulting in small relative velocities and collisional energies that, together with "sticky" microphysical processes (like van der Waals and electromagnetic forces), result in their efficient collisional growth into cm-sized bodies (Weidenschilling \citeyear{1977MNRAS.180...57W}, \citeyear{1980Icar...44..172W}; Birnstiel et al. \citeyear{2011A&A...525A..11B}); these pebbles eventually grow into km-sized planetesimals (see Figure \ref{Kokubo}) and the processes by which these can grow into planetary embryos and planets, via collisions and gravitational interactions, is fairly well understood (Armitage \citeyear{2010apf..book.....A}, and references therein).  

However, the intermediate stage by which cm-sized pebbles grown into km-sized planetesimals poses several challenges, known as the meter-sized barrier. As the cm-sized particles grow and become less coupled to the gas, their relative velocities and collisional energies increase, resulting in collisions that, rather than leading to efficient growth, lead to inefficient sticking, bouncing, or fragmentation (Zsom et al. \citeyear{2010A&A...513A..57Z}); in addition, because the particles are still coupled to the gas, they experience gas drag that results in short inward drift timescales, limiting significantly their lifetime in the disk  and, consequently, their opportunity to grow to sizes unaffected by gas drag (Weidenschilling \citeyear{1977MNRAS.180...57W}; Birnstiel et al. \citeyear{2012A&A...539A.148B}).   

Giant planets can form if the planetesimal growth proceeds fast enough for the embryos to accrete gas from the protoplanetary disk before it dissipates. In the case of the solar system, it is thought that before 10 Myr after the Sun was formed, while the Sun was still embedded in its maternal stellar cluster, and before the gas in the primordial protoplanetary disk dispersed, Jupiter and Saturn formed and scattered planetesimals in the Jupiter--Saturn region to large distances; a fraction of this material had their perihelion lifted beyond the influence of the giant planets due to external perturbations by the stars and the gas in the star cluster, populating the Oort could; but most of the scattered material (75--85\%; Brasser et al. \citeyear{2006Icar..184...59B}) was ejected into interstellar space. In systems where  giant planets have not formed, the smaller-mass planets may not clear their feeding zone and continue to collide on longer timescales.

After the gas is gone, giant planet formation comes to an end but the collisional growth of other planets may continue (Morbidelli et al. \citeyear{2012AREPS..40..251M}). At this point, a process that might have started in an orderly fashion quickly transitions into a fairly chaotic state. The swarms of planetesimals will interact with the growing planets and this can cause their migration and trigger episodes of dynamical instability and orbit readjustment increasing the rate of collisions (Gomes et al. \citeyear{2005Natur.435..466G}; see Nesvorn{\'y} \citeyear{2018ARA&A..56..137N} for a review). 

During these gravitational instabilities, planetesimals can be scattered in or scattered out. In the former case, they can become a source of volatiles to the terrestrial planets region (Morbidelli et al. \citeyear{2000M&PS...35.1309M}; Raymond et al. \citeyear{2009Icar..203..644R}) where the collisional growth of planets is still continuing. In the latter case, it can result in the ejection of a significant fraction of planetesimals into the outer planetary system or into interstellar space. In fact, in the solar system, there is evidence that the planetesimal belts were heavily depleted leaving an asteroid and Kuiper belts that contain only a small faction of their original population. Evidence for a massive primordial Kuiper belt is the existence of KBOs larger than 200 km, which formation by pairwise accretion must have required a number density of objects about two orders of magnitude higher than today. Evidence for a massive primordial asteroid belt comes from the minimum mass solar nebula, showing a strong depletion in the AB region unlikely to be primordial.

\begin{figure}
\begin{center}
\includegraphics[width=18cm]{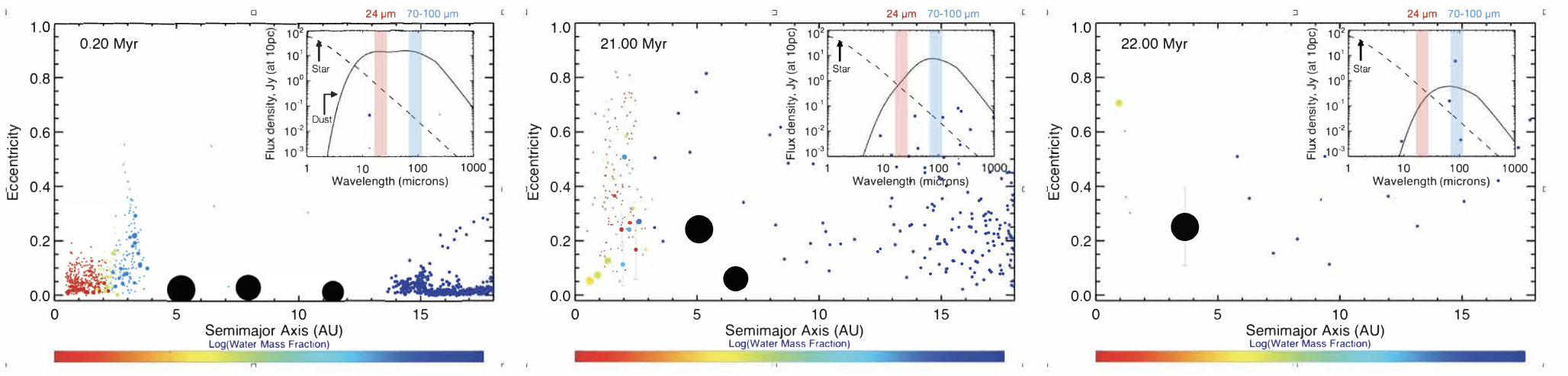}
\end{center}
\caption{Numerical simulations of planet formation by Raymond et al. (\citeyear{2011A&A...530A..62R}, \citeyear{2012A&A...541A..11R}). The planetary system starts with three giant planets and an inner and outer planetesimal belts. The panels show three snapshots of its dynamical evolution, encompassing a gravitational instability that ejects  the majority of the planetesimals. Numerical simulations show that these type of planetesimal-clearing events, of which the Solar system also shows evidence, happen under a wide range of planetary configurations, enriching the interstellar medium with planetesimals. The insert to the top right shows the dust production: the solid line corresponds to the emission from the dust as a function of wavelength and the dashed line to the emission from the star.}
\label{Raymond}
\end{figure}

Even though the efficiency of planetesimal ejection is very sensitive to the planetary architecture and its dynamical history, dynamical models (like those shown in Figure \ref{Raymond}) indicate that planetesimal clearing processes are a natural outcome of the planet-formation processes under a wide range of architectures (Raymond et al. \citeyear{2018MNRAS.476.3031R} and references therein). These processes enrich the interstellar medium with ejected planetesimals. 

\subsection{Debris disks provide observational evidence that planetesimal formation is common and that the planetesimal belts are depleted with time} 
\label{debrisdisks}

The processes described above trigger numerous collisions among planetesimals left in the disk, between planetesimals and the growing planets, or even among  planets, and these collisions produce dust. Dust production also takes place in the outer disk where Pluto-sized objects stir the planetesimal swarms triggering mutual collisions (Kenyon \& Bromley \citeyear{2004AJ....127..513K}). 
 
The timing, duration, and amount of dust released in all these collisional processes vary widely: some of the disk collisional activity is in a pseudo-steady state during a long period (Wyatt et al. \citeyear{2008ARA&A..46..339W}), while other events associated with individual collisions are stochastic and short-lived (Beichman et al. \citeyear{2005ApJ...626.1061B}; Meng et al. \citeyear{2014Sci...345.1032M}; Su et al. \citeyear{2019AJ....157..202S}).  The properties of the dust also differ very significantly: some of the colliding bodies might have formed in situ, while others might have originated in different regions of the disk, inside or outside the different icelines, leading to different parent-body compositions; some of the colliding bodies might be pristine, while others might have been processed; in addition, some of the collisions will be very energetic, altering the composition of the debris, while others will be of low energy, with the composition of the debris dust tracing the parent bodies. 

The interpretation of the dust observations is therefore complex and we are still learning to unveil the clues that are hidden in the dust of planetary systems in the making. But we know that these type of collisions are the origin of the circumstellar dust observed around stars older than a few Myr. This because once the gas of the protoplanetary disk is gone, the primordial dust is subject to more energetic collisions and to Poynting-Robertson drag, and both processes limit the lifetime of the dust particles to the order of 0.01--1Myr. Poynting-Robertson drag is a relativistic effect that results from the interaction of the dust particle with the stellar radiations and can be intuitively understood because in the reference frame of the particle, the stellar radiation appears to come at a small angle forward from the radial direction (due to the aberration of light) that results in a force with a component against the direction of motion; in the reference frame of the star, the radiation appears to come from the radial direction, but the particle reemits more momentum into the forward direction due to the photons blueshifted by the Doppler
effect, resulting in a drag force (Burns et al. \citeyear{1979Icar...40....1B}). This means that the circumstellar dust observed around stars older than a few Myr old is not primordial,  i.e., from the cloud of gas and dust where the star was born, but a debris dust that is replenished as a result of ongoing dust production. 

This dust is critically important to assess whether planetesimal formation is a common process because we cannot directly observe extrasolar planetesimals, like we do in the solar system, but when extrasolar planetesimals collide they produce dust that can have a collective surface area large enough to allow its detection and characterization, shedding light on the underlying planetesimals population. Debris disks are therefore evidence that planetesimal formation is taking place in other systems (see Moro-Mart{\'{\i}}n \citeyear{2013pss3.book..431M} and references therein for a review). 

\begin{figure}
\begin{center}
\includegraphics[width=12cm]{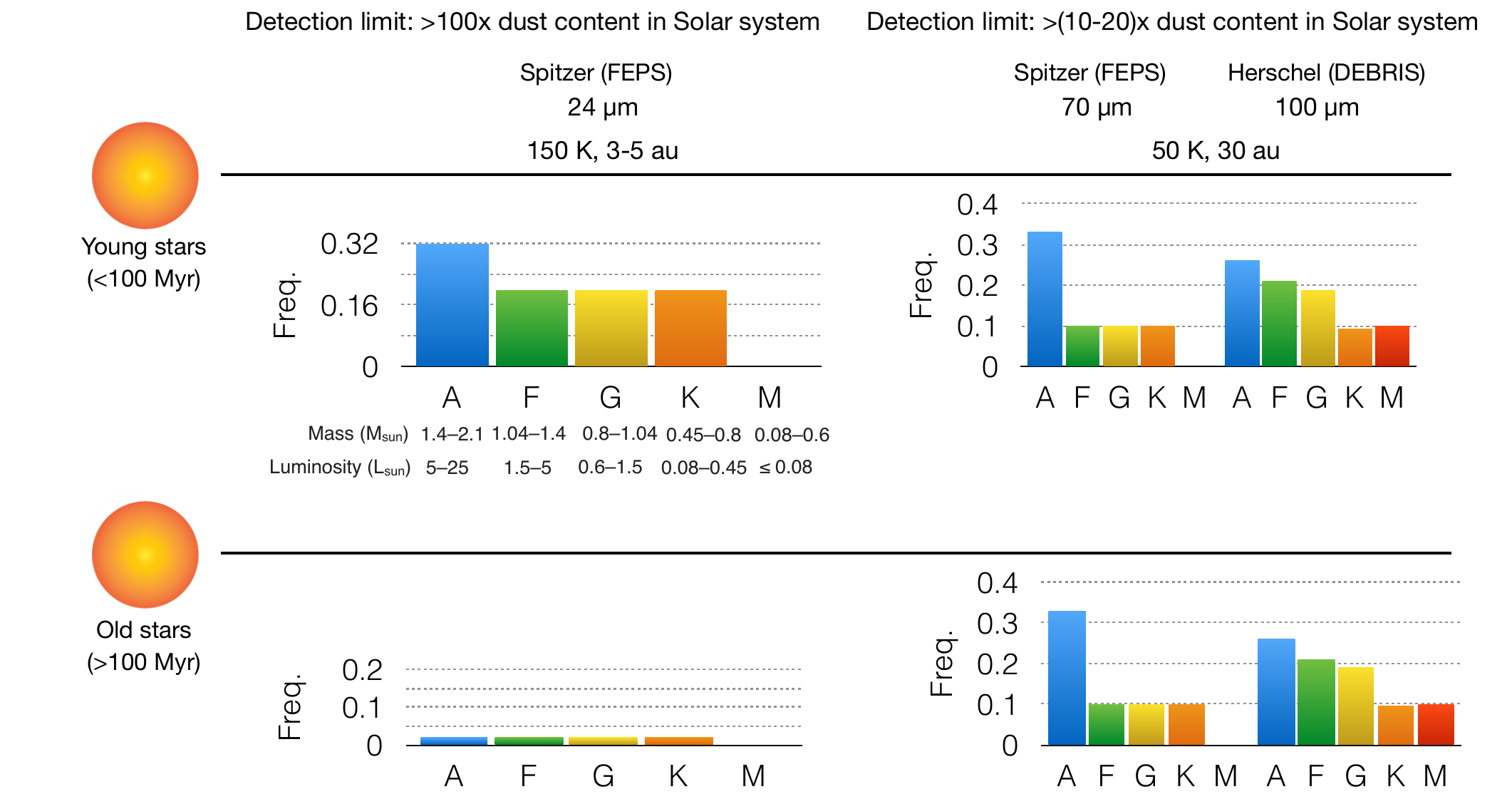}
\end{center}
\caption{Debris disk frequency derived from {\it Spitzer} and {\it Herschel} debris disks surveys for different stellar types.   The mass and luminosity ranges corresponding to the different stellar types are shown below the top, left histogram. The diagrams correspond to: (left) warm dust emission at 24 $\mu$m; (right) cold dust emission at 70 $\mu$m and 100 $\mu$m; (top) young stars with ages $<$100 Myr; (bottom) stars with ages $>$ 100 Myr. The debris disk frequencies are based on results from Meyer et al. (\citeyear{2008ApJ...673L.181M}), Su et al. (\citeyear{2006ApJ...653..675S}), Hillenbrand et al. (\citeyear{2008ApJ...677..630H}), Carpenter et al. (\citeyear {2009ApJS..181..197C}), Moro-Mart{\'\i}n et al. (\citeyear{2015ApJ...801..143M}), Kennedy et al. (\citeyear{2018MNRAS.476.4584K}). 
Debris disks are found around stars with a wide range of metalicities and luminosities, in single and binary systems; because debris disks  are evidence of planetesimals, this indicates that  planetesimal formation is a robust process that can take place under a wide range of conditions.}
\label{DDfreq}
\end{figure}

Figure \ref{DDfreq} shows the frequency of debris disks around stars of different stellar types, derived from {\it Spitzer} and {\it Herschel} debris disk surveys. At 24 $\mu$m, the surveys are sensitive to warm dust at approximately 150 K, that for a solar-type star would be the temperature of dust particles located at 3--5 AU, a distance similar to that of the asteroid belt. At 70-100 $\mu$m, the surveys are sensitive to cold dust at around 50 K that corresponds to a distance of 30 AU, similar to the Kuiper belt.  It is important to note that these surveys are limited by sensitivity. At 24 $\mu$m, the surveys are only able to detect warm dust in systems that contain more than 100 times the amount of warm dust in the solar system. While at 70 $\mu$m and 100 $\mu$m, the detection limit is 10-20 times the amount of cold dust in the solar system. This means that these surveys might only have been able to detect  tip of the iceberg. These surveys indicate that planetesimal discs exist around stars with luminosities that differ by several orders of magnitude; also around stars with wide range of metallicities and with and without binary companions. And it is from these findings that we can infer that debris disks not only indicate that planetesimal formation is taking place in other systems, but that planetesimal formation is a robust process that can take place under a wide range of conditions. 

\begin{figure}
\begin{center}
\includegraphics[width=12cm]{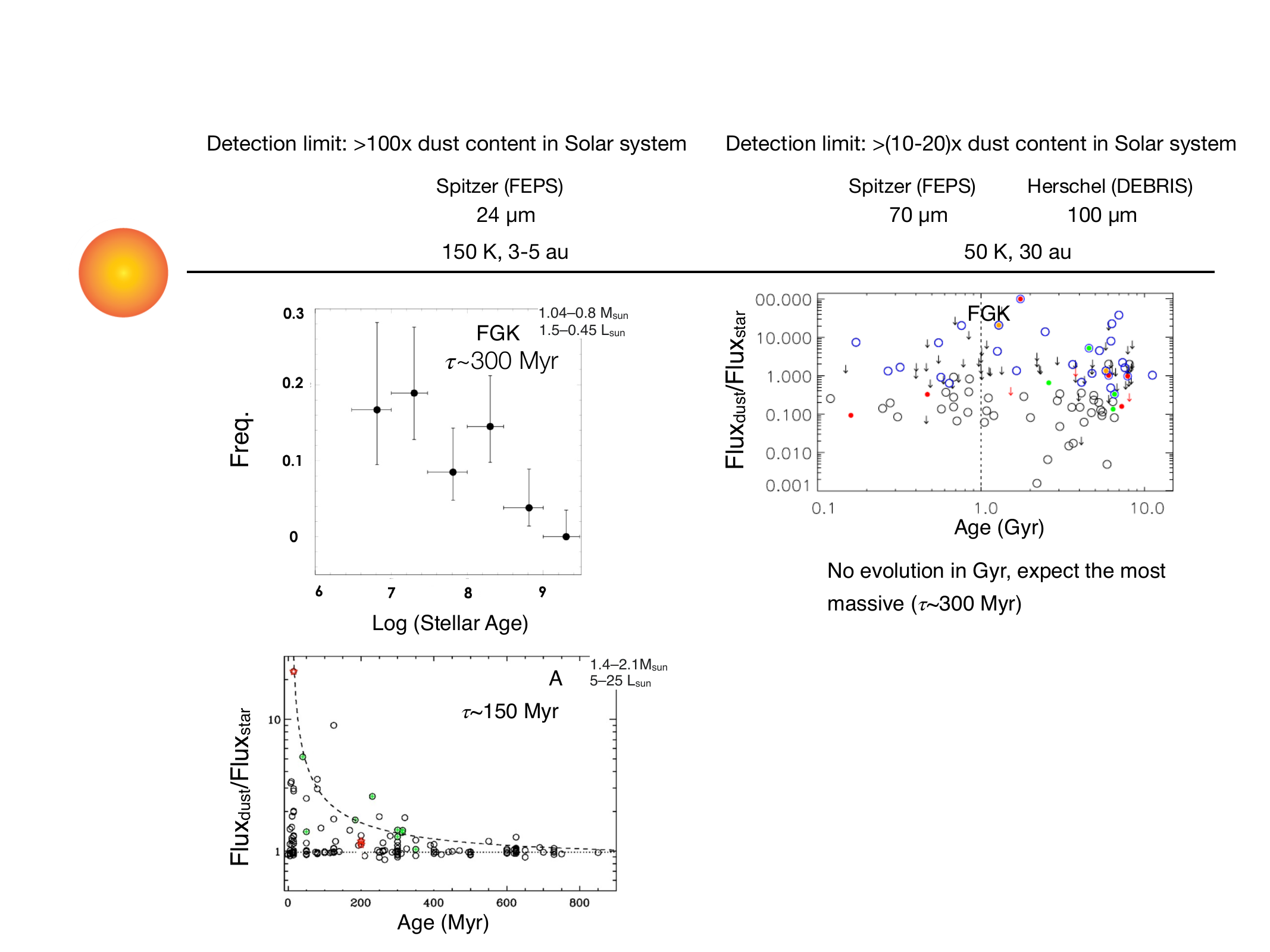}
\end{center}
\caption{Debris disk frequency and dust-to-star flux ratio as a function of stellar age derived from {\it Spitzer} and {\it Herschel} debris disks surveys. The diagrams correspond to: (left) warm dust emission at 24 $\mu$m; (right) cold dust emission at 70 $\mu$m and 100 $\mu$m; (top) FGK-type stars; (bottom) A-type stars. Based on results from Meyer et al. (\citeyear{2008ApJ...673L.181M}), Su et al. (\citeyear{2006ApJ...653..675S}), Moro-Mart{\'\i}n et al. (\citeyear{2015ApJ...801..143M}). Dust production decays with time due to the erosion of the dust-producing planetesimals, more notable in the inner region where the dynamical times are shorter. The dispersion of the bottom panel maybe due to large stochastic collisions .}
\label{DDevolution}
\end{figure}


The {\it Spitzer} and {\it Herschel} debris disks surveys also allow to study how the debris disk frequency and the dust-to-star flux ratio depend on stellar age and, given that we cannot stare at a given system for millions of years, this is a proxy of how the dust-production rate evolves (see Wyatt et al. \citeyear{2008ARA&A..46..339W} for a review). From the left panels on Figure \ref{DDevolution}, we can infer that the dust-producing planetesimals that are located closer to the stars (and produce dust that emits at 24 $\mu$m) disappear on a much shorter timescale than the planetesimals in the outer disk producing dust that can be observed at longer wavelengths (right panel). The reason why it is more common to find cold dust than warm dust around mature stars is because the dynamical times in the inner region of the disk are shorter than in the outer region and this makes planetesimals in the inner region collide more frequently and erode more rapidly, causing the production of warm dust to decay as 1/t. Another interesting aspect that is observed is that over the 1/t envelope there is a lot of dispersion, indicating that dust production in large stochastic collisions may have played an important role in the early evolution of the planetary systems (see bottom left panel from Su et al. (\citeyear{2006ApJ...653..675S}). The cold dust, on the other hand, shows no significant evolution on Gyr times scales. 

The wavy size distributions of the asteroids are a fingerprint that this collisional activity played an important role in the early solar system history, in addition to the depletion of the planetesimal belts due to dynamical ejection discussed earlier (Bottke et al. \citeyear{2005Icar..175..111B}, \citeyear{2005Icar..179...63B}). These studies use the current size distribution of the asteroid belt, together with other observational constraints and collisional evolution models, to calculate the size distribution in the ``primordial'' asteroid. They found that it would have been established early on as a result of a period of collisional activity before Jupiter formed (few Myr), and a period of collisional activity triggered by the planetary embryos (10--100 Myr).  Similarly, the size distribution of the small objects in the Kuiper belt can help constrain its primordial size distribution (Schlichting et al. \citeyear{2013AJ....146...36S}). 

To summarize, the presence of debris disks indicate that planetesimal formation is common and can take place under a wide range of conditions. The discovery of thousands of extra-solar planetary systems is evidence that, in some cases, this has led to the formation of planets in a wide range of planetary architectures (Winn \& Fabrycky \citeyear{2015ARA&A..53..409W}), and dynamical models have shown that the dynamical history of these planetary systems generally involve planetesimal-clearing events. This has led to the idea that these ejected planetesimals, that would be predominantly icy because the majority would have formed outside the snowline in their parent systems, are a component of the interstellar medium. 

\subsection {The unbinding of of exo-Oort cloud objects enrich the interstellar medium with planetesimals} 

In the solar system, the Oort cloud is thought to have formed due to the interplay of planetary scattering and external forces: the forming giant planets scattered the planetesimals in this region out to large distances where they were subject to external influences, like the slowly changing gravitational potential of the cluster, the Galactic tides, and the stellar flybys, with different models favoring different perturbers.  These external influences would have caused the perihelion distances of the scattered planetesimals to be lifted to distances $>>$ 10 au,  where the planetesimals were no longer subject to further scattering events but were also safe from complete ejection and thus remained weakly bound to the solar system, forming the Oort cloud (see for example Brasser et al. \citeyear{2012Icar..217....1B}). Some authors argue that the Oort cloud formed while the Sun was in its birth cluster. Under this scenario, the main perturbers would be the stars and gas in the cluster. These models, however, fail to account for the circularization of the orbits due to the cluster gas (that would impede the comets to be scattered out into the Oort cloud, Brasser et al. \citeyear{2010A&A...516A..72B}), and for the stripping of the outer parts of the Oort cloud ($\ge$ 3 $ \cdot~$~10$^{4}$) by the cluster gravitational potential and neighboring stars.  To account for these caveats, other authors argued that the Oort cloud formed during the late dynamical instability of the solar system, about 0.5 Gyr after it formed. The caveat of these latter models is that this process is not sufficiently efficient (by an order of magnitude) to account for the estimated number of bodies in the Oort cloud (derived from the flux of long-period comets) based on the estimated mass in planetesimals that would have remained $\sim$0.5 Gyr after the solar system formed, i.e. after most of the protoplanetary disk was dispersed (Brasser et al. \citeyear{2010A&A...516A..72B}). 

Even though the formation of the solar system's Oort cloud has still many unknowns, we can expect exo-Oort clouds to form around other stars as the result of the interplay of planetary scattering and external forces that would lead to the lifting of the periastrons of bodies initially orbiting closer to the star (Wyatt et al. \citeyear{2017MNRAS.464.3385W}). Indirect evidence of the presence of a reservoir of comets around other stars are the debris disk systems. There is also evidence that some of these exocomets have been scattered into the inner regions of these system, as suggested by the observation of variable absorption gas features in several of these debris disks (Kiefer et al. \citeyear{2014Natur.514..462K}, Welsh \& Montgomery \citeyear{2015AdAst2015E..26W}), and by the dips in the lightcurve of some Kepler sources (Boyajian et al. \citeyear {2016MNRAS.457.3988B}). 

The reason why we are interested in these exo-Oorts clouds as a potential source of interlopers like 1I/'Oumuamua is because dynamical models show that, over the lifetime of their parent stars, these weakly bound objects are subjected to ejection due to Galactic tides, post-main sequence mass loss, and encounters with other stars or with giant molecular clouds (Veras et al. \citeyear{2011MNRAS.417.2104V},  \citeyear{2012MNRAS.422.1648V}, \citeyear{2014MNRAS.437.1127V}). For  example, for the solar system, Hanse et al. (\citeyear{2018MNRAS.473.5432H}) indicated that over the Sun's main sequence, the Oort cloud will lose 25-65\% of its mass due mainly to stellar encounters, with a second stage of Oort cloud clearing to be triggered by the onset of mass loss as the Sun enters the post-main sequence stage (Veras et al. \citeyear{2012MNRAS.422.1648V}). These ejected objects will contribute to the population of free-floating material and this contribution is expected to be more significant in the Galactic bulge than in the disk or the halo of the Galaxy (due to the more frequent stellar encounters in the former), and in the oldest regions than in the youngest regions (due to the timescale associated to the clearing processes, Veras et al. \citeyear{2014MNRAS.437.1127V}). 

The capture by the solar system of one of these ejected exo-Oort cloud objects today is highly unlikely due to their expected high relative velocity with respect to the Sun, but may have been possible when the solar system was still embedded in its maternal birth cluster (Levison et al. \citeyear{2010Sci...329..187L}; Belbruno \citeyear{2012AsBio..12..754B}), 
with the higher transfer efficiencies being enabled by the lower relative stellar velocities, an order of magnitude lower than today. There is therefore the possibility that we have already observed, or will be able to observe, one of these objects captured from the interstellar medium, but its origin beyond the solar system will likely remain uncertain. 

 \section{Size distribution of interstellar planetesimals}
\label{sizedist}
Needless to say that with only two interstellar objects detected it is not possible to constrain the size distribution of the population. Some of the studies that will be mentioned below adopt a mono-size or an equilibrium size distribution $n(r) \propto r^{-3.5}$ (Dohnanyi \citeyear{1969JGR....74.2531D}).  The only small-body population that can be studied in any detail is that of the solar system. Assuming initially that the source of the interstellar objects are planetesimal disks, it makes sense that the range of possible distributions should encompass that of the small body population in the early solar system, that can be inferred from observations and models. However, it is a challenge for the latter to reproduce simultaneously the observed slopes for the large and small objects and the break radius because this requires to take into account the full dynamical history of the solar system. It is also the case that other planetary systems will likely have experienced a wide range of dynamical and collisional histories and, as a consequence, the size distribution of their ejected bodies will  depend significantly on the degree of dynamical/collisional evolution at the time of the ejection. This is why in the calculations of the number density of interstellar objects that will be discussed below, instead of the equilibrium size distribution $n(r) \propto r^{-3.5}$,  informed by theoretical coagulation and accretion models, we consider a wide range of size distributions characterized by a broken power law of indexes $q_1 = 2-3.5$ (in the small size end), $q_2 = 3-5$ (in the large size end), break radius $r_b$ =  3 km--90 km, r$_{\rm min} \approx~{\rm1}~\mu{\rm m}$, and r$_{\rm max} \approx~{\rm1000~km}$. 

\section{Number density of interstellar planetesimals}
\label{numberdensity}

We need to think of interstellar planetesimals (that originate, for example, from planetary systems in the making or from the release of exo-Oort clouds) as another component of the interstellar medium. It was therefore not a surprise that one of these objects would eventually cross paths with the solar system. And because of the velocity of the Sun with respect to the Local Standard of Rest (LSR; $v_{\rm LSR}$ = 16.5 \kms), it would appear as an interstellar interloper on a hyperbolic trajectory, making it clearly distinguishable from other solar system objects. Therefore, the detection of 1I/'Oumuamua, with a clearly hyperbolic orbit (eccentricity $e$ = 1.197, semi-major axis $a$ =  -1.290, perihelion $q$ = 0.254, and inclination $i$ = 122.6) and high pre-encounter velocity (26.22 \kms, with U, V, W = -11.325, -22.384, -7.629 \kms; Mamajek \citeyear{2017RNAAS...1a..21M}) had been been anticipated for decades. In fact, McGlynn \& Chapman (\citeyear{1989ApJ...346L.105M}) had argued that the lack of extrasolar comet detections was actually problematic because, based on the number density of stars and expected contribution to the ejected planetesimal population, the number density of interstellar objects should be high enough to have detected a significant number entering the solar system. Jewitt (\citeyear{2003EM&P...92..465J}) suggested to use PanSTARRS to constrain their number density. Based on current knowledge of star and planet formation, Moro-Mart{\'\i}n et al. (\citeyear{2009ApJ...704..733M})  predicted the number density of interstellar objects in the Galaxy to be so low that the detection of interstellar comets would require the deep survey capability of the Vera Rubin Observatory (LSST), able to detect smaller objects at greater distances. Engelhardt et al. (\citeyear{2017AJ....153..133E}) reached a similar conclusion a few months prior to 1I/'Oumuamua's detection in the much shallower PanSTARRS data. The visitor was expected, but it had arrived too early. 
 
 \subsection{Number density inferred from 1I/'Oumumua's detection}

Several studies were carried out to estimate the inferred number density of interstellar planetesimals from the detection of 1I/'Oumuamua, adopting different estimates for the detection volume, and what this would imply regarding the contribution per star and how this compares to expectations.  These studies generally agree that the inferred number density is higher than expected in the context of a range of plausible origins.  
Some of the inferred number densities were the following: 0.1 au$^{-3}$ = 8$\cdot10^{14}$ pc$^{-3}$ (Jewitt et al. \citeyear{2017ApJ...850L..36J}, Fraser et al. \citeyear{2018NatAs...2..383F}), 0.012--0.087 au$^{-3}$ = 1--7$\cdot10^{14}$ pc$^{-3}$ (Portegies-Zwart et al. \citeyear{2018MNRAS.479L..17P}), 0.012 au$^{-3}$ = 1$\cdot10^{14}$ pc$^{-3}$ (Gaidos et al. \citeyear{2017RNAAS...1a..13G}), and $<$ 0.006 au$^{-3}$ = 4.8$\cdot10^{13}$ pc$^{-3}$ (lower limit from Feng \& Jones \citeyear{2018ApJ...852L..27F}). These estimates assumed a range of survey times (e.g. 1--2 years Jewitt et al. \citeyear{2017ApJ...850L..36J}; 5 years Portegeis-Zwart et al. \citeyear{2018MNRAS.479L..17P}; 7 years Gaidos et al. \citeyear{2017RNAAS...1a..13G}; 20 years Feng \& Jones \citeyear{2018ApJ...852L..27F}) and also a small range of dark albedos and absolute magnitude that result in a range of average object radius (55 m Jewitt et al. \citeyear{2017ApJ...850L..36J}; 60 m Fraser et al. \citeyear{2018NatAs...2..383F}; 100 m Portegeis-Zwart et al. \citeyear{2018MNRAS.479L..17P}; 115 m Gaidos et al. \citeyear{2017RNAAS...1a..13G}; 50 m Feng \& Jones \citeyear{2018ApJ...852L..27F}). 

We now describe the cumulative number density estimate from Do et al. (\citeyear{2018ApJ...855L..10D}) because it is the one that makes the most detailed calculation of the PanSTARRS detection volume; it also highlights the assumptions and uncertainties inherent to the calculation. They made the assumption that the objects are isotropically distributed and adopted for 1I/'Oumuamua an absolute magnitude of $H$ = 22.1, a nominal phase function with slope parameter $G$ = 0.15, and a velocity at infinity, $v_{\infty}$ = 26 \kms. For these values, they computed the minimum and maximum distance that PanSTARRS could have seen such an object, and assumed that each observation covers 6 deg$^2$. They then calculated the total survey volume by taking into account the effect of gravitational focusing by the Sun, trailing losses due the the tangential velocity of the object with respect to the Earth, the degradation of the signal by the background noise, that the object can come from any approach direction, and assuming that the detection rate for objects as large as 1I/'Oumuamua is given by one detection in the 3.5 year survey time. Assuming that the cumulative number density of interstellar objects down to the detection size $R$ (taken as 1I/'Oumuamua's size) is the inverse of the survey volume, they estimated a cumulative number density of $N_{\rm r \geqslant R}$ =  0.21 au$^{-3}$ $\sim$ 2 $\cdot~10^{15}$ pc$^{-3}$. They noted that this is an underestimate of at most 40\% because the objects have a cumulative size distribution that falls at larger sizes. On the other hand, they pointed out that the detection process is not 100\% efficient over the full 6 deg$^2$ and given these inefficiencies the detection volume could be 2/3--3/4 of the nominal value, so that the number density could be 4/3--3/2 of their inferred number density.  

Below we compare Do et al. (\citeyear{2018ApJ...855L..10D}) number density of interstellar planetesimals inferred from 1I/'Oumuamua's detection ($N_{\rm r \geqslant R}$ $\sim$ 2 $\cdot~10^{15}$ pc$^{-3}$) to that expected from the ejection of planetesimals from protoplanetary disks and from the release of exo-Oort cloud objects. However, in this comparisons it is important to keep in mind that the "observed" value will remain uncertain until more objects are detected and the population becomes better characterized in both its phase space and size distributions.   

\subsection{Expected contribution from the ejection of planetesimals from protoplanetary disks}

\begin{figure}
\begin{center}
\includegraphics[width=12cm]{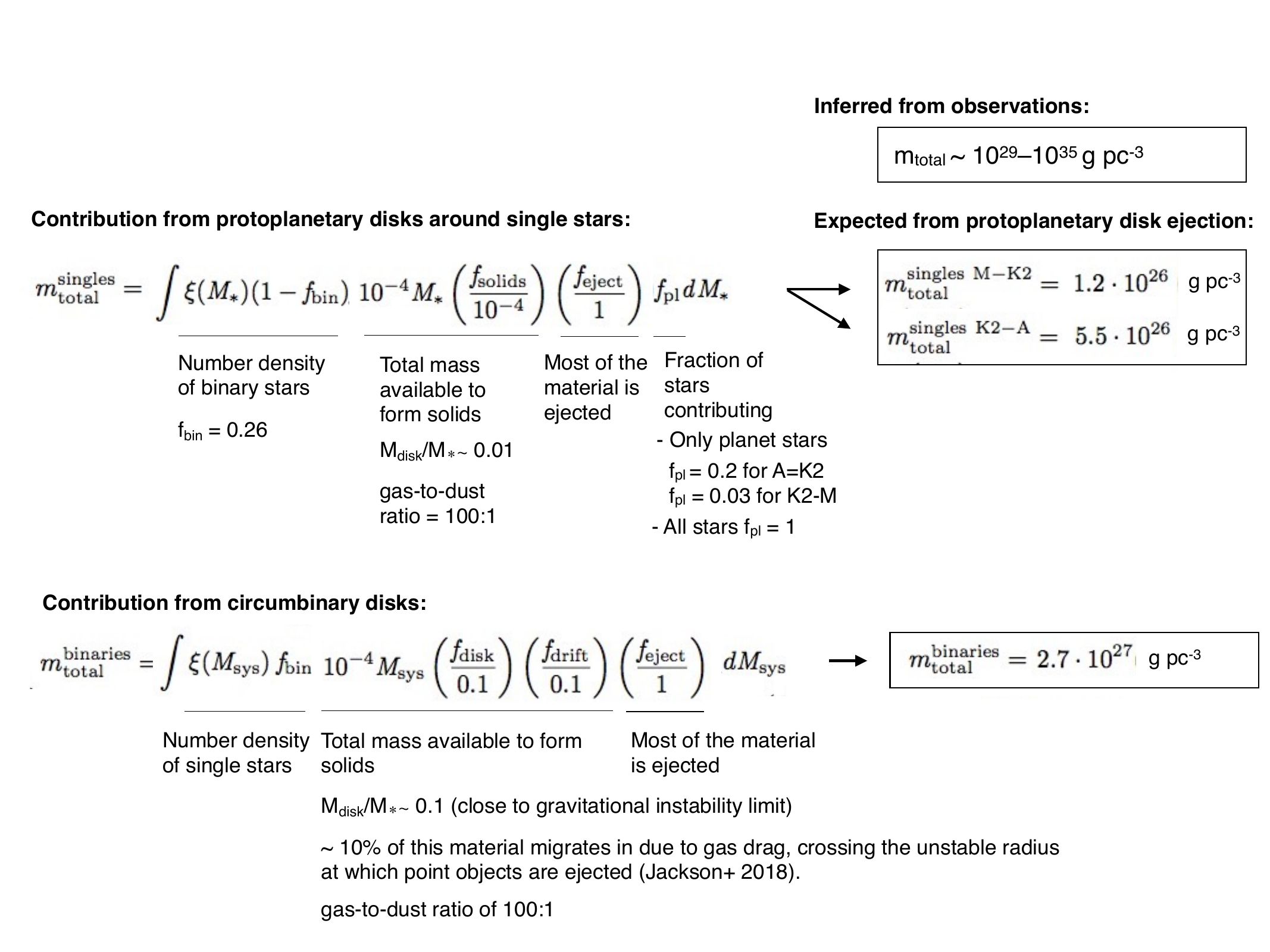}
\end{center}
\caption{Estimate of the mass density of interstellar planetesimals expected from the ejection of planetesimals from protoplanetary disks, compared to that inferred from 1I/'Oumuamua's detection (from Do et al. (2018) and assuming a wide range of possible size distributions for the interstellar planetesimal population. Based on Moro-Mart{\'{\i}}n (\citeyear{2018ApJ...866..131M}). M$_{*}$ is the stellar mass and $\xi(M_*)$ is the number density of stars, from Kroupa et al. (\citeyear{1993MNRAS.262..545K});  $f_{\rm bin}$ is the binary fraction; $f_{\rm solids}$ is the fraction of solid material that is ejected, for which we adopt a value of $10^{-4}$, assuming that the mass of the protoplanetary disks is 1\% of the stellar mass (Andrews \& Williams \citeyear{2007ApJ...671.1800A}), and that 1\% of that material is in solids; $f_{\rm eject}$ is the fraction of the material that is ejected, for which we adopt a value of 1, based on the dynamical models that show how common planetesimal-clearing events are and that only a very small fraction of the planetesimals remain in the disk (see Raymond et al. \citeyear{2018MNRAS.476.3031R} for a review). $f_{\rm pl}$ is the fraction fo stars that contribute, approximated as the fraction of stars known to host the massive planets responsible for the planetesimal ejection (from Marcy et al. \citeyear{2005PThPS.158...24M}; Cumming et al. \citeyear{2008PASP..120..531C}, Johnson et al. \citeyear{2007ApJ...665..785J}).}  
\label{PD}
\end{figure}

We will now describe two population studies  that illustrate the degree of discrepancy between the inferred and expected number density of interstellar planetesimals. The first study, summarized in Figure \ref{PD}, focuses on the contribution to the mass density of interstellar planetesimals from the ejection of planetesimals from protoplanetary disks (Moro-Mart{\'{\i}}n \citeyear{2018ApJ...866..131M}).

For the case of the single stars and close binaries, the contribution is going to depend on the stellar mass, requiring to integrate over all the stellar masses that contribute. For each stellar mass bin, it will depend  on the number density of single stars of that mass and on the total mass available to form solids per star. For the latter, we  adopt $10^{-4}M_{*}$, assuming that the mass of the disk is 1\% that of the stellar mass and 1\% is in the form of solids. Based on the dynamical models mentioned earlier, that show how common planetesimal-clearing events are and that only a small fraction of the planetesimals remain in the disk, we will assume that most of the solids are ejected. This is likely to be an overestimate. Ejection is indeed efficient around 1 M$_{\rm Jupiter}$ planets located at 1--30 au, 0.1--10 M$_{\rm Jupiter}$ planets at $\sim$5 au, Saturn-mass planets at 10-30 au, eccentric planets and long-period giant planets (Wyatt et al. \citeyear{2017MNRAS.464.3385W}), but this is for the material that crosses the orbits of these planets. Because planets are necessary for the planetesimal ejection, we further assume that the stars that are known to host massive planets are the ones that contribute (but we also consider the case that all stars contribute).The majority of the planetesimals expected to be ejected from these sources would be icy. This is because the Safronov number for a planet of a given mass increases with orbital radius, making ejection more efficient beyond the snowline (see e.g. Raymond et al. \citeyear{2018MNRAS.476.3031R}).  However, Raymond et al. (\citeyear{2018ApJ...856L...7R}) point out that some of these objects might be devolatized because of multiple close passages close to its host star before being ejected, as it might have been the case of 1I/'Oumuamua that, as mentioned, did not show any cometary activity (so there is no evidence it was icy). 

For the case of the circumbinary disks, their contribution to the mass density of interstellar planetesimals is calculated in a similar way but in this case the total mass available to form solids per system is  $10^{-4}M_{\rm sys}$,  where $M_{\rm sys}$ is the mass of the binary system. This is based on Jackson et al. (\citeyear{2018MNRAS.477L..85J}) that estimates that the mass of the circumbinary disk is 10\% that of the binary system, and that 10\% of that material migrates due to gas drag and crosses the unstable radius at which point the objects are ejected; we also assume that 1\% of that material is solids. Based on this work also we also assume that all the wide binaries would contribute. This study estimates that 64\% of the ejected material in this case would be devolatized, having spent significant time close to the binary stars before being ejected, while the remaining 36\% would be icy. 

The resulting mass density from the calculations described above are listed at the right side of Figure \ref{PD} under "expected from protoplanetary disk ejection". From single and wide binaries (adding the contribution from all relevant spectral types), we expect a mass density $m_{\rm total}$ $\sim$  7 $\cdot~10^{26}$ g pc$^{-3}$. From circumbinary disks, we expect $m_{\rm total}$ $\sim$  3 $\cdot~10^{27}$. These values are significantly lower than the mass density "inferred from observations" of $m_{\rm total}$ $\sim$ $10^{29}$--$10^{35}$ that is calculated using the number density estimate in Do et al. (\citeyear{2018ApJ...855L..10D}), adopting the range of possible size of size distributions described in Section \ref{sizedist}. These calculations assume for 1I/'Oumuamua a characteristic bulk density of 1.0 g/cm$^3$, consistent with its tumbling state and composition and an effective radius of 80 m (Drahus et al. \citeyear{2018NatAs...2..407D}).

\subsection{Expected contribution from the release of planetesimals from exo-Oort clouds}

\begin{figure}
\begin{center}
\includegraphics[width=12cm]{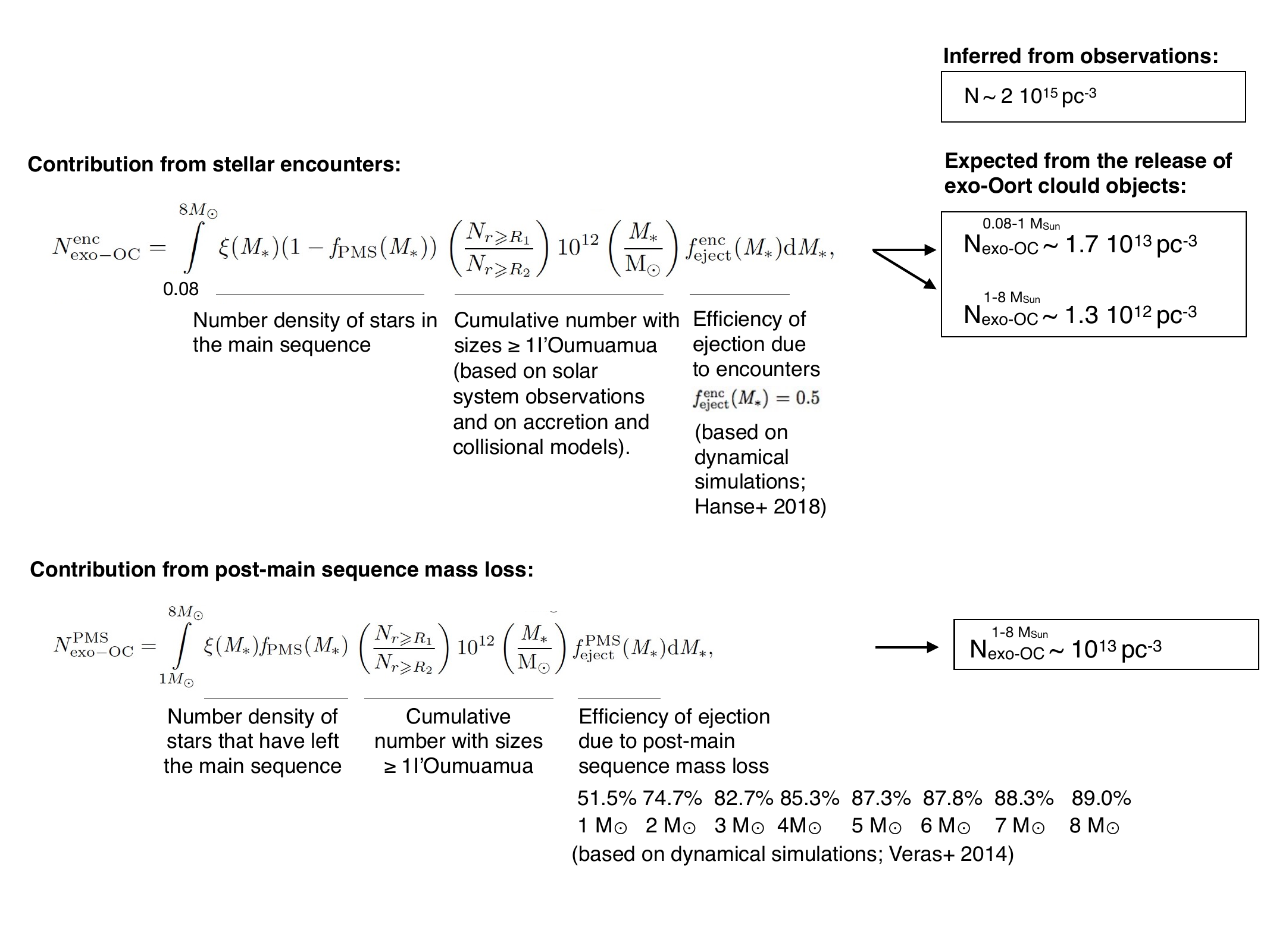}
\end{center}
\caption{Estimate of the number density of interstellar planetesimals expected from the release of exo-Oorts objects due to stellar encounters and post-main sequence mass loss, compared to the number density inferred from 1I/'Oumuamua's detection (from Do et al. (2018). Based on Moro-Mart{\'\i}n (\citeyear{2019AJ....157...86M}).  M$_{*}$ is the stellar mass and $\xi(M_*)$ is the number density of stars, from Kroupa et al. (\citeyear{1993MNRAS.262..545K});  $f_{\rm PMS}({\it M_{*}})$ is the fraction of stars that have reached the post-main sequence stage, assuming a constant star formation rate over the age of the Galactic disk. $\left({N_{r \geqslant R_1}\over N_{r \geqslant R_2}}\right)\rm 10^{12} \left({{\it M_{*}} \over {M_{\odot}}}\right)$ is the cumulative number of bodies with sizes equal or larger than 1I/'Oumuamua's, where $R_1$ = 0.08 km (1I/'Oumuamua's adopted effective radius) and $R_2$ = 1.15 km; here, we are assuming that for a given exo-Oort cloud, the number of bodies with diameter larger than 2.3 km is similar to that of the solar system's Oort cloud, approximated to be 10$^{12}$, scaled to the mass of the parent star.  To calculate $\left({N_{r \geqslant R_1}\over N_{r \geqslant R_2}}\right)$, we assume that the size distribution can be approximated as the broken power-law, based on solar system observations and on accretion and collisional models. ${\it f_{\rm eject}^{\rm PMS}}({\it M_{*}})$ and ${\it f_{\rm eject}^{\rm enc}}({\it M_{*}})$ is the efficiency of ejection of exo-OC bodies due to post-main sequence mass loss and stellar encounters, derived from dynamical models in Hanse et al. (\citeyear{2018MNRAS.473.5432H}), and Veras et al. (\citeyear{2014MNRAS.437.1127V}), respectively.}
\label{OC}
\end{figure}

The second population study (from Moro-Mart{\'\i}n \citeyear{2019AJ....157...86M}) is summarized in Figure \ref{OC}. It focuses on the contribution to the number density of interstellar planetesimals from the release of exo-Oort cloud objects due to mass loss, close encounter with other stars or the galactic tide. We assume that the Oort cloud contains $\sim$ 10$^{12}$ objects with diameters $\geqslant$ 2.3 km. This estimate is based on the  flux of long-period comets, thought to be launch from the Oort cloud into the inner solar system due to perturbations by the Galactic tide, or by encounters with stars or giant molecular clouds, together with subsequent perturbations by the planets (Brasser \& Morbidelli \citeyear{2013Icar..225...40B} and references therein).  Following Hanse et al. (\citeyear{2018MNRAS.473.5432H}), for a given exo-Oort  cloud we assume that its number of bodies larger than 2.3 km is similar to that of the solar system's Oort cloud scaled to the mass of the parent star, in our case 10$^{12} \left({{\it M_{*}} \over {M_{\odot}}}\right)$. We further assume that all stars contribute, irrespective of whether or not they are planet hosts. This latter assumption likely makes our back-of-the-envelope estimate an upper limit because, even though the models of Brasser et al. (\citeyear{2010A&A...516A..72B}) found that the Oort cloud formation efficiency is similar at a wide range of Galactocentric distances, Wyatt et al. (\citeyear{2017MNRAS.464.3385W}) shows that the parameter space (in terms of planetary architecture) to form an Oort cloud is quite restricted and in the solar system is populated by Uranus and Neptune. 

 Based on long-period comet observations, it is estimated that the solar system's Oort cloud has an isotropic distribution with perihelion $q$ $\gtrsim$ 32 au and inner and outer semimajor axes of $a_{\rm min}^{\rm OC}$ $\sim$ 3 $ \cdot~$~10$^{3}$ and $a_{\rm max}^{\rm OC}$ $\sim$ 10$^{5}$ au, respectively. Following Hanse et al. (\citeyear{2018MNRAS.473.5432H}), we scale the inner and outer edge of a given exo-Oort cloud to the Hill radius of its parent star in the Galactic potential. 
 
The two processes expected to dominate the exo-Oort clouds clearing are: (1) post-main sequence mass loss, for the stars in the 1--8 M$_{\odot}$ mass range that have reached this stage of stellar evolution; and (2) stellar encounters, for the stars that are still on their main sequence. Their ejection efficiencies depend on where these objects are located and are based on dynamical models in the literature.  

The resulting number density from the calculations described above are listed at right side of Figure \ref{OC} under "expected from the release of exo-Oort cloud objects". The number density of interstellar planetesimals expected to be triggered by stellar encounters is $\sim$ 3$\cdot~10^{13}$ pc$^{-3}$ and from post-main sequence mass loss (of stars in the 1--8 M$_{\odot}$ mass range) is $\sim 10^{13}$ pc$^{-3}$. Again, these values are significantly lower than the number density "inferred from observations" of $N_{\rm r \geqslant R}$ =  0.21 au$^{-3}$ $\sim$ 2$ \cdot~10^{15}$ pc$^{-3}$ (from Do et al. \citeyear{2018ApJ...855L..10D}). 

In more intuitive units this means that if we assume that 1I/'Oumuamua is representative of a population that is uniformly distributed,  from its detection we can estimate that there are about 10,000 interstellar objects like 1I/'Oumuamua within the orbit of Neptune at any given time, that is a sphere of about 30 AU in radius, while from the ejection of planetesimals from protoplanetary disks or from the release of planetesimals from exo-Oort  one would expect up to 100. There are many uncertainties in these calculations but this discrepancy points out that we still have a lot to learn about the population of interstellar planetesimals and their origin. 

\subsection{Proposed solutions for the discrepancy between the inferred and expected number density of interstellar planetesimals}

\subsubsection{1I/'Oumuamua could have originated in a young nearby system}

One of the proposed solutions to address the discrepancy between the inferred and the expected number density of interstellar planetesimals is that 1I/'Oumuamua is not representative of an isotropic distribution of interstellar planetesimals, which could have led to an overestimate of the background density. This would be the case, for example, if it is originating from a nearby young planetary system, as suggested by Gaidos et al. (\citeyear{2017RNAAS...1a..13G},  \citeyear{2018MNRAS.477.5692G}) . In this case, the ejected population would likely have a highly anisotropic distribution, resulting in large fluctuations in space density.  

There are several observations that support a young age for 1I/'Oumuamua. One is the color of the object, found not to be as red as the ultra-red bodies in the outer solar system (Jewitt et al. \citeyear{2017ApJ...850L..36J}), thought to be reddened by space weathering (from cosmic rays and ISM plasma); some authors argue that this suggests that 1I/'Oumuamua has not been exposed to space weathering for Gyr (Gaidos et al. \citeyear{2017RNAAS...1a..13G}; Feng \& Jones \citeyear{2018ApJ...852L..27F}; Fitzsimmons et al. \citeyear{2018NatAs...2..133F}).  The second observation is 1I/'Oumuamua's entering velocity, found to be within 3--10 \kms~of the velocity of the {\it LSR} (Gaidos et al. \citeyear{2017RNAAS...1a..13G}; Mamajek \citeyear{2017RNAAS...1a..21M}, Do et al. \citeyear{2018ApJ...855L..10D}); the fact that this velocity is similar to that of many young stellar associations also supports that 1I/'Oumuamua has not traveled in interstellar space during Gyr because otherwise  dynamical heating (due to passing stars, clouds, spiral arms, and star clusters) would have increased its relative velocity with respect to the {\it LSR}.  Gaidos et al. (\citeyear{2017RNAAS...1a..13G}) estimate an age of $\ll$1 Gyr, while Feng \& Jones (\citeyear{2018ApJ...852L..27F}) indicate that the probability of observing the object with a velocity $<$10 \kms~with respect to the {\it LSR} is 0.5, 0.26, and 0.13, for ages of 0.1 Gyr, 1 Gyr, and 10 Gyr, respectively. It has also been suggested that 1I/'Oumuamua is tumbling, likely due to a collision in its parent system, and that because the damping time for an object with its properties would be $\sim$ 1 Gyr (due to stresses and strains that result from its complex rotation), this implies that the object is younger than that age (Drahus et al. \citeyear{2018NatAs...2..407D}). 

Some authors have attempted to identify the star or association from which 1I/'Oumuamua originated, but these studies generally do not take into account the errors in stellar positions, which is the most important source of uncertainty (Dybczy{\'n}ski \& Kr{\'o}likowska \citeyear{2018A&A...610L..11D}). An  study using GAIA data  traced back the trajectory of 1I/'Oumuamua and of about 7 million stars and identified four stars with trajectories that may have intersected the trajectory of 1I/'Oumuamua (Bailer-Jones et al. \citeyear{2018AJ....156..205B}), but as the authors point out 1I/'Oumuamua may have passed within 1 pc of about 20 stars and brown dwarfs every Myr. Given the typical interstellar distance between encounters, large perpendicular displacements can be produced, making it difficult to predict the result of successive stellar encounters (Zhang \citeyear{2018ApJ...852L..13Z}).  The uncertainty regarding its pre-perihelion trajectory (because it is not known when  1I/'Oumuamua's non-gravitational acceleration, discussed below, appeared)  further complicates the search of its birth system.

\subsubsection{1I/'Oumuamua could be the result of a formation/fragmentation process with a narrow size distribution}

Another of the proposed solutions to address the discrepancy between the inferred and the expected background density is that 1I/'Oumuamua is the result of a fragmentation process with a narrow size distribution, which would have led to an overestimate of the background density. The adopted range of size distributions (described in Section \ref{sizedist}) is based on observations of the small body population in the solar system and  theoretical coagulation and accretion models, but the size distribution of the initial population of building blocks remains as one of the major open questions in planet formation. Streaming instability, for example, may lead to a narrow size distribution (Johansen \& Lambrechts \citeyear{2017AREPS..45..359J}) but these models are not yet able to predict it at the small end (including 1I/'Oumuamua's size). 

As mentioned in Section \ref{discovery}, 1I/'Oumuamua's brightness variation has been interpreted as evidence that its shape had an unusual elongated morphology and this has led to the proposal that it might be a fragment of a tidally disrupted planetesimal. If 1I/'Oumuamua originated in a tidal disruption event, and this event resulted in a narrow size distribution, this might help resolve the tension between the inferred and expected number density of planetesimals, as the former would have been overestimated.  

Raymond et al. (\citeyear{2018ApJ...856L...7R}, \citeyear{2018MNRAS.476.3031R}) suggested that 1I/'Oumuamua is a fragment of a planetesimal that was tidally disrupted when it passed within the tidal disruption radius of a gas giant, arguing that 100 m might be the characteristic size of the fragments, as opposed to a wider size distribution.  Their dynamical simulations show that indeed $\sim$1\% of planetesimals pass within the tidal disruption radius of a gas giant on their pathway to ejection.  Zhang \& Lin (\citeyear{2020NatAs...4..852Z}) suggested 1I/'Oumuamua is an elongated fragment that resulted from the tidal disruption of a planet, or a small body, that came too close to its parent star and that was later ejected to interstellar space. 

It has also been proposed that 1I/'Oumuamua originated in a tidal disruption event around a white dwarf system, instead of a nascent planetary system. This is motivated by the observations that indicate that some white dwarfs show infrared excess emission due to debris dust, thought to be produced by colliding or disrupting planetesimals, and that some white dwarfs show atmospheric pollution, thought to be produced by planetesimals that accrete onto the evolved star. This has led to the idea that 1I/'Oumuamua may have been ejected from one of these systems, either by direct ejection or as a result of a tidal disruption event that could have involved a binary system, and that the collisional fragmentation could have channel most of the original material into a narrow size distribution (0.1--1 km; Rafikov \citeyear{2018ApJ...861...35R}; {\'C}uk \citeyear{2018ApJ...852L..15C}; Hansen \& Zuckerman \citeyear{2017RNAAS...1a..55H}). A caveat of this evolved star scenarios is that the expected population of remnants would exhibit kinematic characteristics similar to that of old stars, that would have experienced dynamical heating in the galaxy by gravitational scattering with massive objects, and this contrary to the observed kinematic properties of 1I/'Oumuamua.  

A critical aspect that is still missing from all the tidal disruption scenarios mentioned above are the models that predict that the tidal fragments are indeed elongated and that their size distribution is narrow.

\section{An unexpected trajectory leading to unconventional ideas about origin}

\begin{figure}
\begin{center}
\includegraphics[width=8cm]{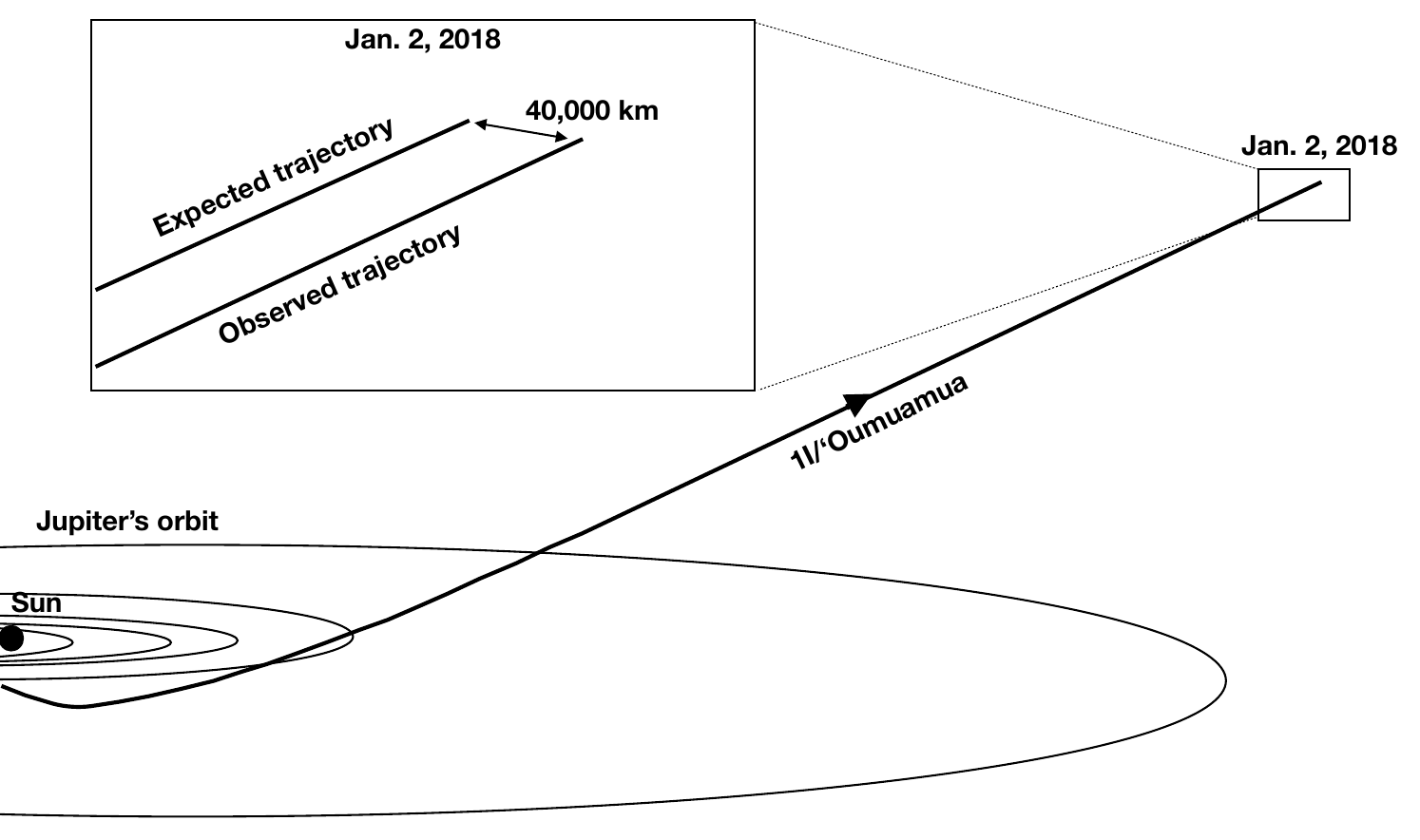}
\end{center}
\caption{Effect of non-gravitational acceleration on 1I/'Oumuamua's trajectory.}
\label{1Iacceleration}
\end{figure}

Upon close inspection of 1I/'Oumuamua's outgoing orbit (see Figure \ref{1Iacceleration}), it was discovered that the object was experiencing a non-gravitational acceleration, $\Delta a = a_0 \left( r \over {\rm AU} \right)^n$, with the best fit for $n$ = -2 and $a_0 = (4.92 \pm 0.16) \times 10^{-4}~{\rm cm~s}^{-2}$ (Micheli et al. \citeyear{2018Natur.559..223M}). These type of jet-like force is typical in comets and is caused by the mass loss that happens on the dayside of the nucleus, where the ice is sublimating and therefore it was suggested to be the cause of 1I/'Oumuamua's excess acceleration (Micheli et al. \citeyear{2018Natur.559..223M}). The big caveat with this interpretation is that 1I/'Oumuamua never showed any evidence of the gas and dust loss observed in comets (Jewitt et al. \citeyear{2017ApJ...850L..36J}, Meech et al. \citeyear{2017Natur.552..378M}). Observations by {\it Spitzer} imply 3-$\sigma$ upper limit to CO outgassing that is four orders of magnitude lower than that invoked by Micheli et al. (\citeyear{2018Natur.559..223M}) to account for the observed non-gravitational acceleration (Trilling et al. (\citeyear{2018AJ....156..261T}). If one assumes that CO and CO$_2$  have a similar relative abundance with respect to H$_2$O as found in comets (CO + CO$_2$ being $\sim$ 15\% that of H$_2$O), even when adopting outgassing levels of CO and CO$_2$ at the 3-$\sigma$ upper limit, the inferred level of H$_2$O outgassing would  be 1\% of that required, implying that for this scenario to work, the object would have to be devolatized of CO and CO$_2$ prior to {\it Spitzer} observations (Trilling et al. \citeyear{2018AJ....156..261T}), a plausible solution given that they have lower volatilization temperatures than H$_2$O. Another explanation is that 1I/'Oumuamua had a chemical composition different from that characteristic of small bodies in the solar system due to different formation conditions. 'Oumuamua ISSI Team et al. (\citeyear{2019NatAs...3..594O})  point out that the CO and CO$_2$ ratios in comets has recently been found to be far greater than was previously known. Another constraint to the H$_2$O outgassing level comes from the lack of CN emission (Ye et al. \citeyear{2017ApJ...851L...5Y}); in this case, the required level of H$_2$O outgassing would require 1I/'Oumuamua to be depleted in CN by a factor of 15 compared to typical comet abundances; only two comets in the solar system show this type of depletion ('Oumuamua ISSI Team et al. \citeyear{2019NatAs...3..594O}). 

The absence of dust loss is  problematic because the gas generally drags small dust particles along but no dust loss was found either; it has been argued that maybe only large dust grains were dragged along, in which case the dust outflow could have been unnoticed because no observations were sensitive to large dust grains. 'Oumuamua ISSI Team et al. (\citeyear{2019NatAs...3..594O}) point out that a few long-period comets preferentially eject large particles due to an unknown mechanism that is currently not understood, but this is very very unusual (David Jewitt, private communication). 

Rafikov (\citeyear{2018ApJ...867L..17R}) pointed out that another challenge to the outgassing scenario is the expectation that the implied outgassing torques would have spun-up the object in a timescale of few days, leading to its breakup. However, Seligman et al. (\citeyear{2019ApJ...876L..26S}) reported that an elongated object with outgassing at the subsolar could account for the observed light curve amplitude and period  without a disrupted spin-up. 
  
Seligman \& Laughlin (\citeyear{2020ApJ...896L...8S}) have suggested that 1I/'Oumuamua's acceleration was indeed due to outgassing from a new type of body made of molecular hydrogen ice, a cosmic hydrogen iceberg that originated in the starless coldest regions of a molecular cloud, but the temperature to keep hydrogen on ice form is close to the ambient cosmic microwave background, making the survival of such an object unlikely. Alternately, Jackson \& Desch (\citeyear{2021JGRE..12606706J}) have suggested that 1I/'Oumuamua was made of N$_2$ ice, a fragment from an exo-Pluto surface, the problem in this case is to account for the inferred number density (Siraj \& Loeb \citeyear{2021arXiv210314032S}).

Given the challenges to the outgassing scenario, Bialy \& Loeb (\citeyear{2018ApJ...868L...1B}) put forward the suggestion that this excess acceleration could be due to radiation pressure, $a = {PA \over m} = \left({L\odot \over 4 \pi r^2 c}\right) \left({A \over m}\right) C_R = 4.6 \times 10^{-5} \left({r \over {\rm AU}} \right)^{-2} \left({m/A \over {\rm g~cm^{-2}}}\right)^{-1} C_R ~{\rm cm~s}^{-2}$, where $A$ and $m$ are the area and mass of the object, respectively, $C_R$ is of order unity and depends on the objects composition and geometry and $r$ is the distance to the Sun. This was motivated because it has the same $r^{-2}$ radial dependency as that found by Micheli et al. (\citeyear{2018Natur.559..223M}) that best fits 1I/'Oumuamua outgoing trajectory, $\Delta a = (4.92 \pm 0.16) \left( r \over {\rm AU} \right)^{-2}$. Equating the two expression we have that ${A \over m} = {1 \over (9.3 \pm 0.3) \times 10^{-2}~C_R} ~{\rm cm^{2} g^{-1}}$, and this led Bialy \& Loeb (\citeyear{2018ApJ...868L...1B}) to conclude that I/'Oumuamua has the morphologies of a thin sheet 0.3--0.9 mm in width, representing a new class of  interstellar material of an unknown natural or artificial origin (like a lightsail). 'Oumuamua ISSI Team et al. (\citeyear{2019NatAs...3..594O}) point out that such a planar geometry is inconsistent with the shape and amplitude of 1I/'Oumuamua's light-curve.

\subsection{Cosmic dust bunnies and primordial planet building blocks}
\label{fluffysec}

\begin{figure}
\begin{center}
\includegraphics[width=8cm]{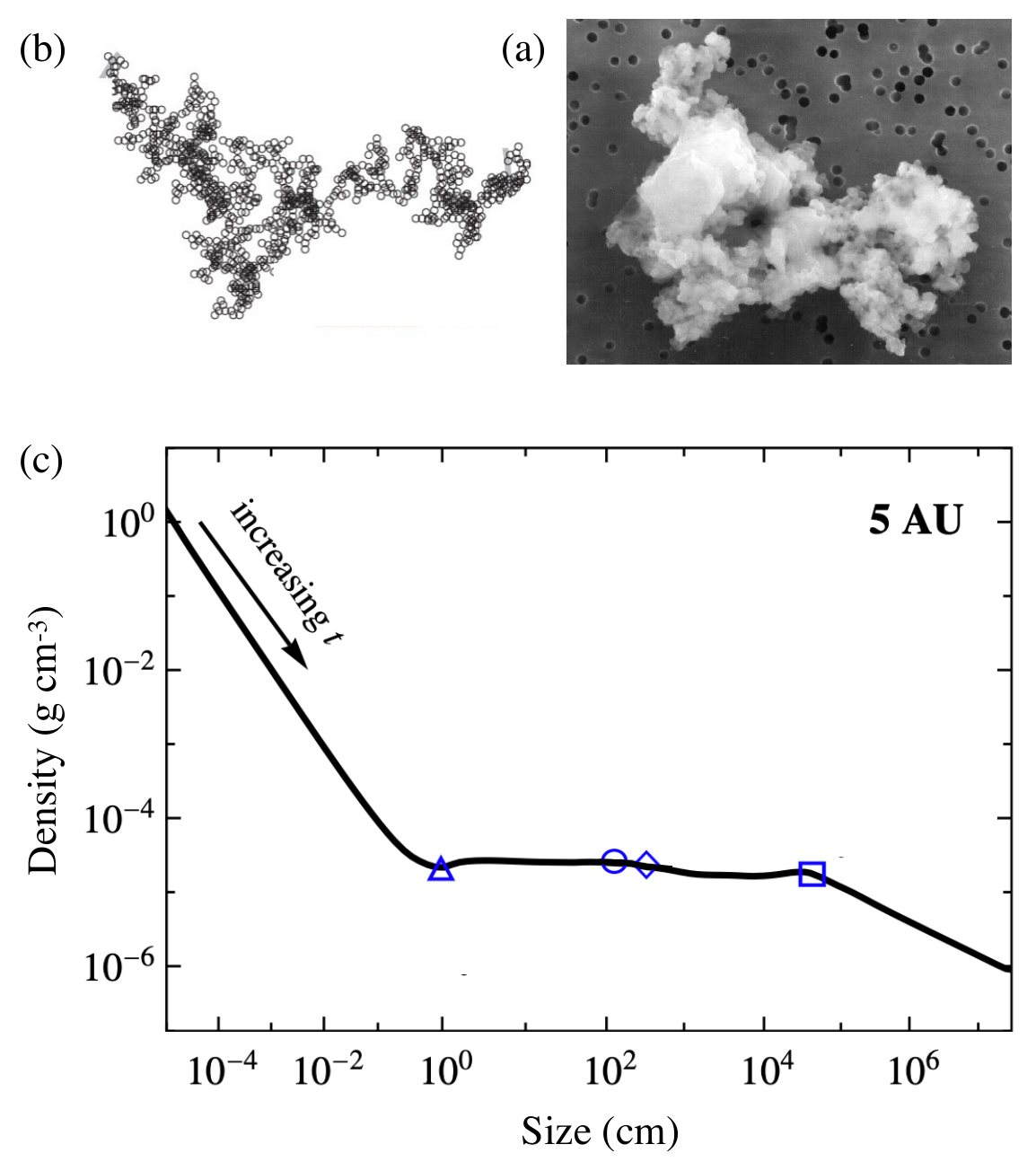}
\end{center}
\caption{(a) Interplanetary dust particle collected in space (credit: D. E. Brownlee). (b)  Results from numerical modeling of the growth of dust aggregates (Okuzumi et al. \citeyear{2009ApJ...707.1247O}). (c) Results from numerical modeling of the growth of planetesimals via dust collisions in a protoplanetary disk. These models show that beyond the iceline, the planetesimals will have less and less density as they grow (Okuzumi et al. \citeyear{2012ApJ...752..106O}).}
\label{fluffy}
\end{figure}

As an alternative to this planar sheet scenario, and given the strict {\it Spitzer} upper limits to outgassing, Moro-Mart{\'\i}n (\citeyear{2019ApJ...872L..32M}) suggested that if 1I/'Oumuamua were to have a mass fractal structure with a high area-to-mass ratio, radiation pressure could be responsible for its excess acceleration. Fractal structures are found in many forms of nature and are thought to arise because their formation processes involve an element of stochasticity, like particle collisions in a solution, in a turbulent circumstellar cloud, or in a protoplanetary disk, environments characterized by low local particle concentrations and large diffusion lengths. Interplanetary dust particles, like the one shown at the top, right of Figure \ref{fluffy} show a fractal structure with a mass fractal dimension of ${N_f}$ = 1.75 (Katyal et al. \citeyear{2014JQSRT.146..290K}; ${N_f} \sim$ 3 would be due to a compact, dense structure, and ${N_f} \sim$ 1 to "stringy" one). The area-to-mass ratio of a mass fractal is given by ${A \over m} = {D^2 \over \left({D \over D_0}\right)^{N_f} \rho_0 D_0^3} = \left({D_0 \over D}\right)^{N_f-2} {1 \over \rho_0 D_0}$ and the bulk density by $\rho = {m \over V } = {\left({D \over D_0}\right)^{N_f} \rho_0 D_0^3 \over D^3}= \left({D \over D_0}\right)^{N_f-3}\rho_0$, where $D$ is the aggregate size, $D_0$ is the primary particle size (assuming a single-size distribution),  and $\rho_0$ is the primary particle bulk density (Moro-Mart{\'\i}n \citeyear{2019ApJ...872L..32M}). Equating the area-to-mass ratio expected from a mass fractal to the value that would be required to support 1I/'Oumuamua's radiation pressure scenario leads to a bulk density of O($10^{-5}$) g cm$^{-3}$ for 1I/'Oumuamua's size. This would imply that 1I/'Oumuamua is like a cosmic dust bunny or snowflake with a bulk density about 100 times less than air. Given that the lowest density solid known is graphene aerogel, about 10 times less dense than air, and it is synthetically produced, the first outstanding question is whether such an ultra-low density aggregate could form naturally and survive. 

At the microscopic level, there is observational evidence of the existence of fluffy aggregates with extremely low densities $<$ 10$^{-3}$ g cm$^{-3}$ detected by the GIADA instrument on ROSETTA and possibly also by Stardust (Fulle et al. \citeyear{2015ApJ...802L..12F}). Additional evidence comes from experimental studies (Blum \& Schr{\"a}pler \citeyear{2004PhRvL..93k5503B}) and from numerical simulations of grain growth that have investigated the porosity evolution of dust aggregates as they grow to planetesimal sizes (see Kataoka \citeyear{2017ASSL..445..143K} for a review). At first, their collisional energies are not high enough to restructure the colliding aggregates and their porosity rapidly increases as they grow. Collisional compression starts when the collisional energy exceeds the rolling energy of the aggregate (required to roll a primary particle over a quarter of the circumference of another primary particle in contact) but it is inefficient and the porosity of the aggregate still continues to increase as it grows because most of the colliding energy is spent compressing the new voids that are created when two aggregates collide and stick to each other, rather than compressing the voids that were already present in the colliding aggregates. Suyama et al. (\citeyear{2008ApJ...684.1310S}) investigated the porosity evolution of icy dust aggregates growing in laminar protoplanetary disks similar to the minimum-mass solar nebula via sequential equal-mass, head-on collisions. They found that the collisional compression stage results in fluffy aggregates (like the one shown in Figure \ref{fluffy} top, left) with a fractal dimension of 2.5 and extremely low densities $<$ 10$^{-4}$ g cm$^{-3}$, noting that this density would be even lower if one were to account for oblique collisions that result in elongated aggregates. Okuzumi et al. (\citeyear{2012ApJ...752..106O}) extended this model to study the growth of icy aggregates beyond the snowline of protoplanetary disks and found that the resulting aggregates at the end of the collisional compression stage have even lower densities of $\sim$ 10$^{-5}$ g cm$^{-3}$ for a wide range of aggregate sizes, encompassing 1I/'Oumuamua estimated size (see Figure \ref{fluffy} bottom). There is therefore the interesting possibility that the collisional grow of icy\footnote{Silicate grains, on the other hand, would result in more efficient collisional compression because of their lower surface energy compared to icy grains, resulting in higher densities.} dust particles beyond the snowline of a protoplanetary disk might naturally produce fractal aggregates of the size of 1I/'Oumuamua with bulk densities that are low enough to support, or to contribute significantly, to the radiation pressure-driven scenario. This origin could also account for its unusual physical properties because such a fluffy aggregate would be very different from the more compact solar system objects taken as reference. 

Very little is known about the intermediate products of planet formation because we can only observe the two extremes of the size distribution:  dust on the smallest end, and the planets on the largest end. So the hypothesis that 1I/'Oumuamua could be an one of these intermediate products is extraordinary because it could open a new observational window to study the primordial building blocks of planets around other stars, and this can set unprecedented constraints on planet formation models. For example, numerical models find that fluffy icy aggregates, like the ones discussed above that 1I/'Oumuamua may represent, can accelerate planetesimal growth because of their increased cross-section, helping to avoid several growth barriers (Suyama et al. \citeyear{2008ApJ...684.1310S}). They can overcome the radial drift barrier within 10 AU for a minimum mass solar nebula model, facilitating planetesimal growth in the inner regions of protoplanetary disks, outside the water snowline (Okuzumi et al. \citeyear{2012ApJ...752..106O}). They can also overcome the  fragmentation barrier if they are constituted by primary particles 0.1 $\mu$m in size because the expected maximum collisional velocities in the disk midplane are generally smaller than the fragmentation threshold velocities for these type of aggregates (Kataoka et al. \citeyear{2013A&A...557L...4K}). Finally, fluffy icy aggregates are not subject to the bouncing barrier because of the small  number of primary particles that are in contact with each other. The existence of fluffy aggregates can also have an impact on planet formation because their porosity could delay the onset of runaway growth (as the escape velocity decreases with increasing porosity, Okuzumi et al. \citeyear{2012ApJ...752..106O}).  

Flekk{\o}y et al. (\citeyear{2019ApJ...885L..41F}) have studied whether such an ultra-porous structure could survive the hazards of interstellar travel as it would be subject to rotational and tidal forces during its journey. They found that such an object could survive and that the interaction of an ultra-low density aggregate with the solar radiation could explain the changes observed in the rotational period of 1I/'Oumuamua. What is not clear is if such an ultra-porous structure could survive tidal disruption during ejection and, in the case of 1I/'Oumuamua, the passage near the Sun.  Given the results of Kataoka et al. (\citeyear{2013A&A...557L...4K}) regarding aggregate compression due to ram pressure by the disk gas, another aspect that needs to be studied is to assess the viability of the icy fractal aggregate hypothesis  is whether its extremely low density could be maintained while in the parent system, during its long interstellar journey, and when entering the solar system. 

There are other origins that have been been proposed for an ultra-low density 1I/'Oumuamua. Luu et al. (\citeyear{2020ApJ...900L..22L}) proposed that 1I/'Oumuamua grew from a collection of dust particles in the coma of an active comet that then escaped, while Sekanina (\citeyear{2019arXiv190108704S}) proposed that 1I/'Oumuamua is an ultra-porous devolatilized fragment that resulted from the disintegration of an "ordinary" km-sized extrasolar comet as it passed near the sun. Regarding this latter hypothesis, because the parent extrasolar comet is significantly larger than the  80 m radius assumed for 1I/'Oumuamua, this would increase significantly the discrepancy between the inferred and expected number density of interstellar objects discussed above. 

\section{2I/Borisov: a planetesimal ejected from the cold outer edge of a distant planetary system}

\begin{figure}
\begin{center}
\includegraphics[width=8cm]{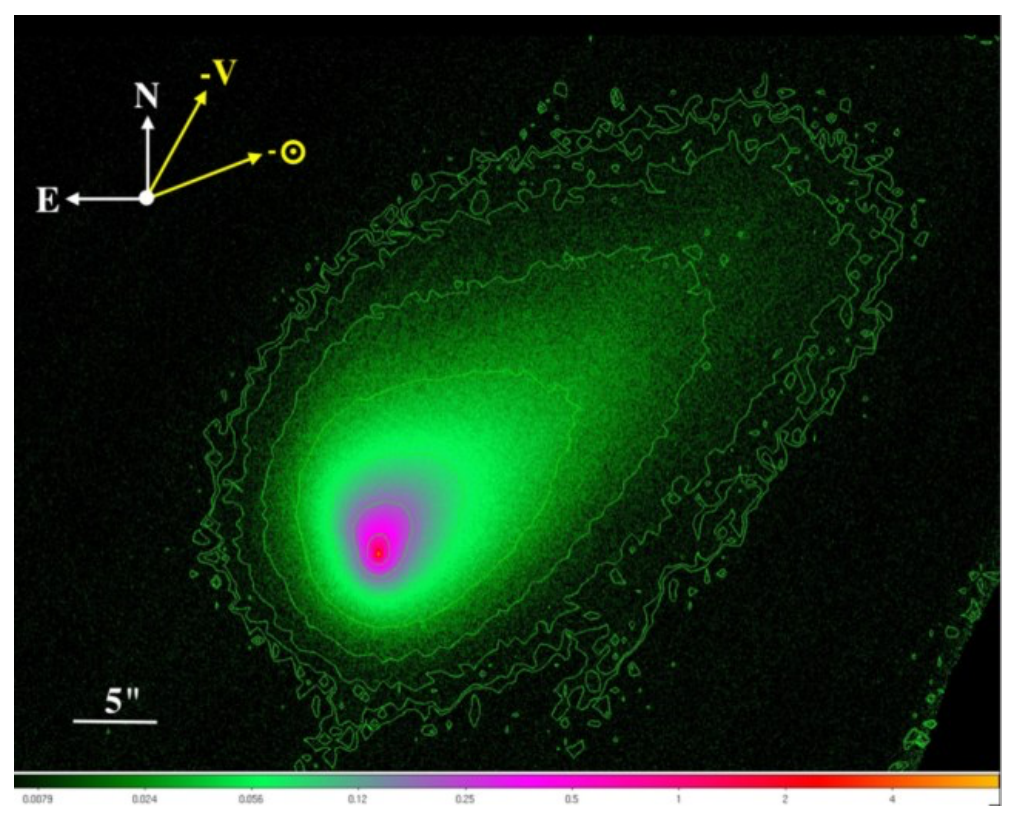}
\end{center}
\caption{2I/Borisov. Figure from Jewitt et al. (\citeyear{2020ApJ...888L..23J}).}
\label{2I}
\end{figure}

2I/Borisov was the second interstellar interloper that was detected two years after 1I/'Oumuamua (Borisov 2019). Even though its hyperbolic trajectory (Figure \ref{2Itrajectory}) did not pass as close to the Earth as 1I/'Oumuamua, it was easier to detect well before perihelion thanks to its larger size and cometary activity. Whereas 1I/'Oumuamua's nature remains a mystery, 2I/Borisov's  unquestionable cometary nature (Jewitt \& Luu \citeyear{2019ApJ...886L..29J}) indicates that it is an ice-rich planetesimal ejected from an extra-solar planetary system, in agreement the the expectation that most ejected planetesimals would have formed outside the snowline in their parent systems. Its coma has made the determination of the size of its nucleus challenging, estimated to be 200 m--500 m, without evidence of being as elongated as 1I/'Oumuamua (Figure \ref{2I}; Jewitt et al. \citeyear{2020ApJ...888L..23J}). It was also observed that its nucleus experienced rotational bursting of one or more meter-size boulders, an event involving negligible mass that would not affect the survival of the object (Jewitt et al. \citeyear{2020ApJ...896L..39J}). 2I/Borisov's trajectory did show an excess acceleration, but in this case it can be accounted for by gas and dust loss. It was found to be very rich in CO (Cordiner et al. \citeyear{2020NatAs...4..861C}), containing significantly more CO than H$_2$O with an abundance more than three times higher than in any comet in the inner solar system (Bodewits et al. \citeyear{2020NatAs...4..867B}). Because CO ice is so easy to vaporize, this probably means that this extrasolar comet originated from the outermost regions of its host planetary system, beyond the CO iceline where the CO can stay in ice form.  

\section{Interstellar planetesimals are potential seeds for planet formation and life}

\subsection{Planetesimal can be transferred between young planetary systems}

\begin{figure}
\begin{center}
\includegraphics[width=8cm]{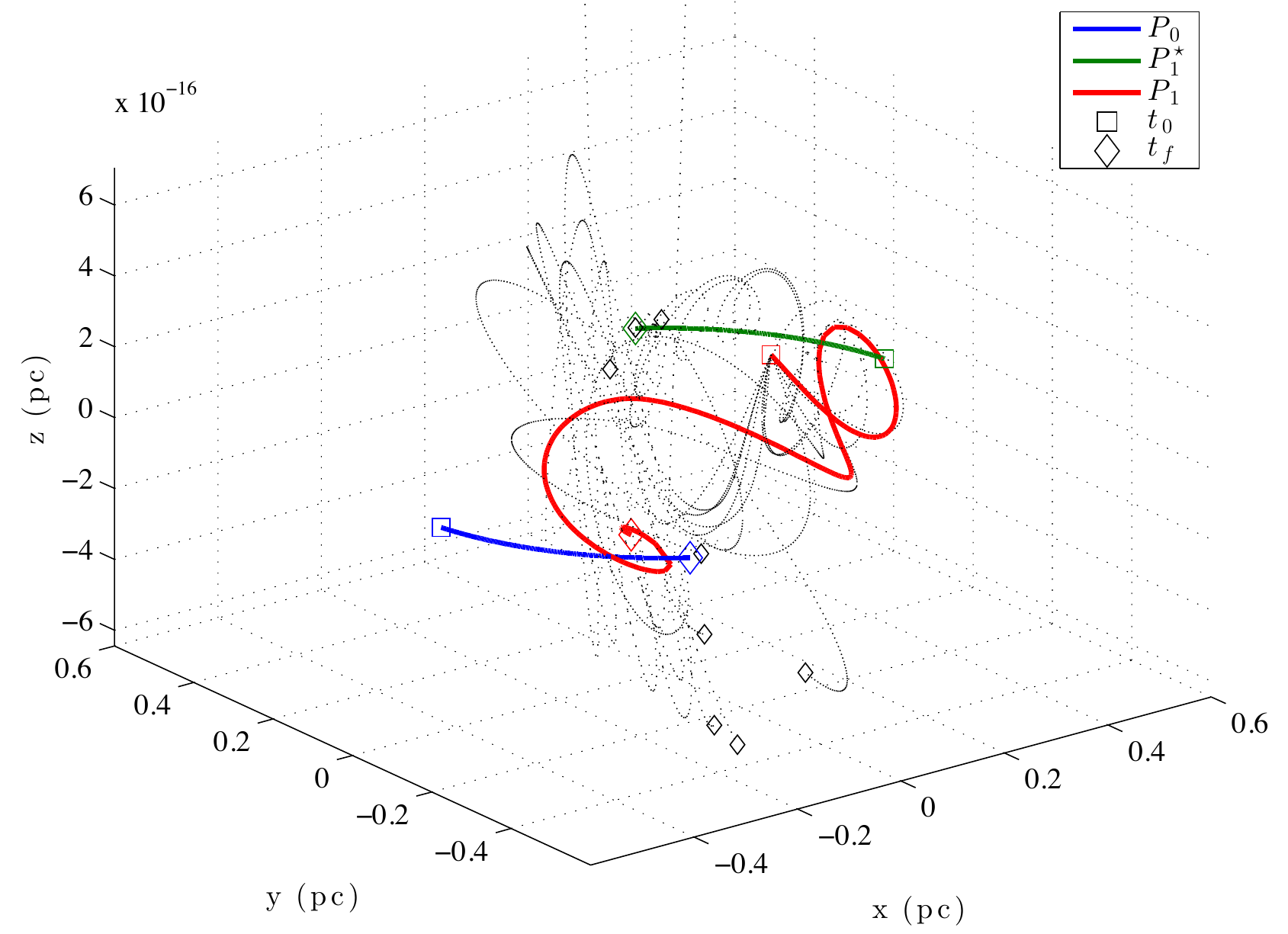}
\end{center}
\caption{Montecarlo simulations of the transfer of planetesimals between two stars in an open cluster using quasi-parabolic, chaotic orbits (dotted black lines). The stars are moving at small relative velocities ($\sim$ 1 \kms). Planetesimals are ejected from a planet-host star.  One of the planetesimals (in red) is captured by another planet-host star. The green and blue lines represent the orbits in the cluster of the parent and target stars, respectively. Based on Belbruno et al. (\citeyear{2012AsBio..12..754B}).}
\label{WT}
\end{figure}

The capture by the solar system of planetesimal ejected from other systems is highly unlikely due to their expected high relative velocity with respect to the Sun. But stars are generally born in clusters, where the stars have relative stellar velocities that are about an order of magnitude lower than in the solar neighborhood today. Belbruno et al. (\citeyear{2012AsBio..12..754B}) found that the probability that planetesimals were transferred  between our solar system and nearby planetary systems while still embedded in the birth cluster is 0.15\%. This is about nine orders of magnitude higher than when using hyperbolic orbits. The increased transfer probability is enabled by the chaotic, quasi-parabolic orbits considered inside the cluster (Figure \ref{WT}), as opposed to hyperbolic orbits like those of 1I/'Oumuamua and 2I/Borisov, where the objects just fly by. 

\begin{figure}
\begin{center}
\includegraphics[width=12cm]{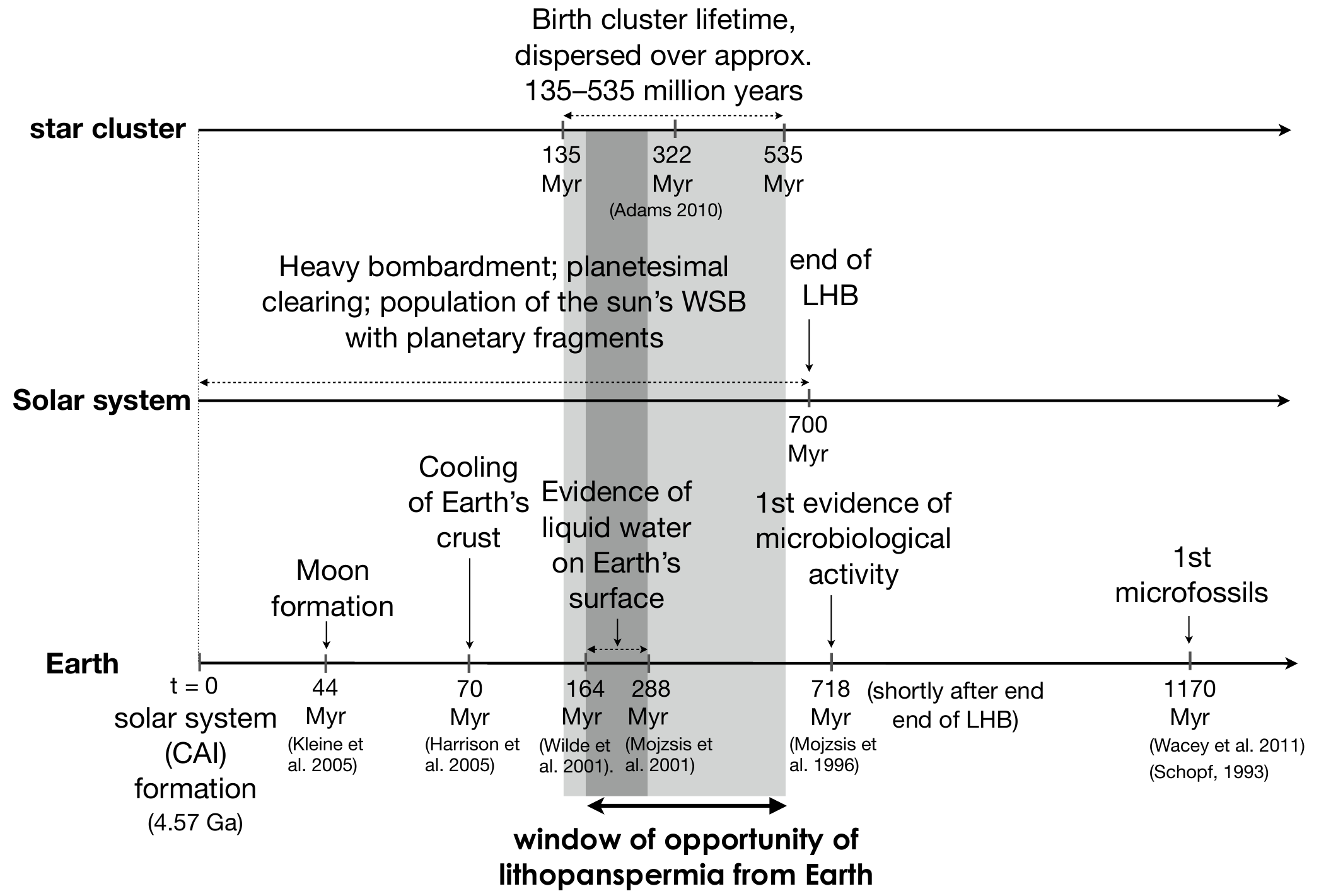}
\end{center}
\caption{Chronological events chart that shows the timeline of events for the Earth, the solar system and the star cluster. A period of massive bombardment and planetesimals clearing in the solar system may have overlapped with the period that the star cluster was still bound. This means that, from the dynamical point of view, a window of opportunity existed for the transfer of material between the solar system and other planetary systems in the cluster. At the time, there is evidence that liquid water could have been present near the Earth's surface. This transfer of solid material is of interest in the context of lithopanspermia if life had an early start in the solar system or in other planetary systems in its neighborhood. Based on Belbruno et al. (\citeyear{2012AsBio..12..754B}).}
\label{WTchart}
\end{figure}

This increased probability of transferring solid material within the birth cluster could have extraordinary and thought-provoking implications in the context of lithopanspermia. Figure \ref{WTchart} is a chronological events chart that shows the timeline of events for the Earth (bottom), the solar system (middle), and the birth cluster (top). At Earth, there is evidence of liquid water near its surface fairly early on, when the Earth was about 200 Myr old.  At the time, there is evidence of a lot of dynamical activity in the solar system, involving planetesimal clearing and heavy bombardment. It is also possible that at that time the stellar cluster where the Sun was born was still bound (its dispersal time is uncertain). The fact that all those factors could have overlapped means that, from the dynamical point of view, a window of opportunity existed for the transfer of solid material between the solar system and other planetary systems in the cluster and this is very interesting in the context of lithopanspermia if life had an early start in the solar system or in other planetary systems in its neighborhood.  Belbruno et al. (\citeyear{2012AsBio..12..754B}) estimated that of the order O(10$^{14}$)--O(10$^{16}$) planetesimals larger than 10 kg could have been transferred between our solar system and one of its neighbors before the birth cluster dispersed. 

\subsection{Interstellar planetesimals can act as condensation nuclei for planet formation}

As mentioned in Section \ref{planetformation}, an unsolved problem in planet formation theory is how cm-sized pebbles grow into km-sized planetesimals, as particles approaching a meter size grow inefficiently due to increased collisional energies; in addition, they have short inward drift timescales due to gas drag that limit significantly their lifetime in the disk and their opportunity to grow to sizes unaffected by gas drag. Several mechanisms have been proposed to alleviate this meter-sized barrier: the effect of the gravitational collapse of the solid component in the disk (Youdin \& Su \citeyear{2002ApJ...580..494Y}; Youdin \& Goodman \citeyear {2005ApJ...620..459Y}) and the formation of over-dense filaments in the disk fragmenting gravitationally into clumps of small particles that collapse and form planetesimals (Johansen et al. \citeyear {2007Natur.448.1022J}); the effect of the velocity distribution of the dust particles that allows for the existence of low-velocity collisions that favor the growth of a small fraction of particles into larger bodies (Windmark \citeyear{2012A&A...544L..16W}); the presence of pressure maxima that can trap dust grains (Johansen et al. \citeyear{2004A&A...417..361J}); the effect of low porosity mentioned in Section \ref{fluffysec} (Okuzumi et al. \citeyear{2009ApJ...707.1247O}) and the electric charging of dust aggregates (Okuzumi \citeyear{2009ApJ...698.1122O}) to favor sticking over fragmentation;  and collisional growth between particles of very different masses colliding at high velocity (Booth et al. \citeyear{2018MNRAS.475..167B}). 

 The discovery of 1I/'Oumuamua and 2I/Borisov gave rise to another proposal:  the "seeding" of the star- and planet-forming environments with interstellar planetesimals. Grishin et al. (\citeyear{2019MNRAS.487.3324G}) studied the scenario in which interstellar objects are captured by the protoplanetary disk due to gas drag, while Pfalzner \& Bannister (\citeyear{2019ApJ...874L..34P}) studied how, during the cloud-formation process, interstellar objects could be incorporated to the molecular cloud in the same way as the interstellar gas and dust, and from there could be incorporated to the the star-forming disks.  Moro-Mart{\'\i}n \& Norman (\citeyear{2021arXiv211015366M}) refined these calculations and found that the number of trapped interstellar objects can be significant. When assuming a background number density of 2$\cdot$10$^{15}$ pc$^{-3}$ (from Do et al. \citeyear{2018ApJ...855L..10D}), a velocity dispersion of 30 \kms (characteristic of the young stars in the Galaxy from where  planetesimals would have originated) and an equilibrium size distribution (with a power-law index of 3.5), the study found that the number of interstellar objects captured by a molecular cloud and expected to be incorporated to each protoplanetary disk during its formation is
6$\cdot$10$^{8}$ (50 cm--5 m), 
2$\cdot$10$^{5}$  (5 m--50 m), 
6$\cdot$10$^{1}$  (50 m--500 m), 
2$\cdot$10$^{-2}$  (500 m--5 km). After the disk formed, the number of interstellar objects that it could capture from the interstellar medium during its lifetime is  
6$\cdot$10$^{11}$ (50 cm--5 m), 
2$\cdot$10$^{8}$ (5 m--50 m), 
6$\cdot$10$^{4}$ (50 m--500 m), 
2$\cdot$10$^{1}$ (500 m--5 km).
The latter estimate assumes a field environment. 

In an open cluster environment, where the relative velocities of the stars are small and the exchange of planetesimals between them can be more effective, the study showed that if 1\% of the clusters stars have undergone planet formation, the number of interstellar objects that a neighboring protoplanetary disk in the cluster could capture during the disk lifetime is a factor of $\sim$600 larger than the field values quoted above.  

The interest of these trapped interstellar objects is that they could have sizes large enough to overcome the drift barrier, as the maximum drift speed in the protoplanetary disk is found for bodies around a meter in size, decreasing significantly for larger sizes: for planetesimals with bulk densities of 3 g$\cdot$cm$^{-3}$ located at a radial distance of 5 AU in a solar nebula, Weidenschilling (\citeyear{1977MNRAS.180...57W}) found drift timescales of O(10$^{2})$ yr, O(10$^{4})$ yr, O(10$^{5})$ yr, and O(10$^{7})$ yr, for particles with sizes of 1 m, 10 m, 100 m, and 1 km, respectively. The settling timescale into the disk mid-plane also decreases with size (Weidenschilling \citeyear{1980Icar...44..172W}; see e.g. Figure 2 in Bate \& Lor{\'e}n-Aguilar \citeyear{2017MNRAS.465.1089B}). Both processes can favor the rapid growth of these trapped planetesimals into larger bodies via the direct accretion of the sub-cm sized dust grains in the protoplanetary disk, with their more rapid settling time allowing them to get a boost in accretion growth as they settle into the mid-plane. The important conclusion is that planet formation, in particular in star clusters, can be significantly influenced by stellar capture of interstellar planetesimals, so there may be cluster-wide environmental effects.  

The estimated numbers of trapped objects listed  above correspond to an equilibrium size distribution. When considering the wide range of possible size distributions for interstellar objects discussed in Section \ref{sizedist}, and the expected background density of interstellar objects discussed in Section \ref{numberdensity}, the corresponding results for the expected number of trapped objects show a wide range of possible values. We will be able to narrow down these estimates as the population of interstellar objects becomes better characterized, both observationally and via simulations, in particular its background number density, size, and velocity distributions. 

Future simulations of molecular clouds and star-forming environments (Pfalzner et al. \citeyear{2021arXiv210608580P}) and of planet formation should take into account the presence of a population of trapped interstellar objects. The latter simulations are necessary to estimate how many of these condensation nuclei would be necessary to account for the planets and small body populations in solar and extra-solar planetary systems and to assess if the trapped interstellar objects can play an important role.
In the solar system, for example, we have to account for the formation of a large small-body population of asteroids, comets and Kuiper belt objects; the Kuiper belt alone is thought to host of the order of 10$^5$ objects $>$ 100 km, and long-period comet observations indicate that of the order of 10$^{12}$ objects $>$ few km populate the Oort cloud. Planet-formation models are necessary to assess how many interstellar seeds should be sufficient to account for the large dynamical range of masses.

This trapping mechanism could not provide the first generation of "seeds" but, because it will have an increasingly important role with time, as the number density of interstellar planetesimals in the Galaxy increases, it might alleviate the potential problem of requiring all the planet-forming disks to have fine-tuned conditions to overcome the meter-sized barrier.

\section{Future prospects}

Our knowledge of the interstellar object population is in its infancy. Based on the detection of 1I/'Oumuamua, it is estimated that, at any given time, $\sim$ 10,000 interstellar planetesimals $\gtrsim$ 100 m are crossing the solar system within the orbit of Neptune, implying a population of $\sim10^{24}$--$10^{25}$ in the Galaxy. Yet, only two objects have been identified so far and they are remarkably different, the unusual properties of 1I/'Oumuamua contrasting sharply with the unquestionably extrasolar cometary nature of 2I/Borisov.  Starting in 2023, the {\it Vera Rubin Observatory} will systematically survey the sky more deeply than has ever been attempted and it will do it repeatedly, revolutionizing this field with the detection of dozens to a hundred of new interstellar objects (Levine et al. \citeyear{2021arXiv210811194L}), in addition to a tenfold increase in the number of solar system object detections (Ivezi{\'c} et al. \citeyear{2019ApJ...873..111I}). These observations will constrain the number density, size, and velocity distributions of interstellar objects and will allow their early monitoring with follow-up observations, including with {\it JWST}, allowing to study their nature and origin. 

\begin{figure}
\begin{center}
\includegraphics[width=8cm]{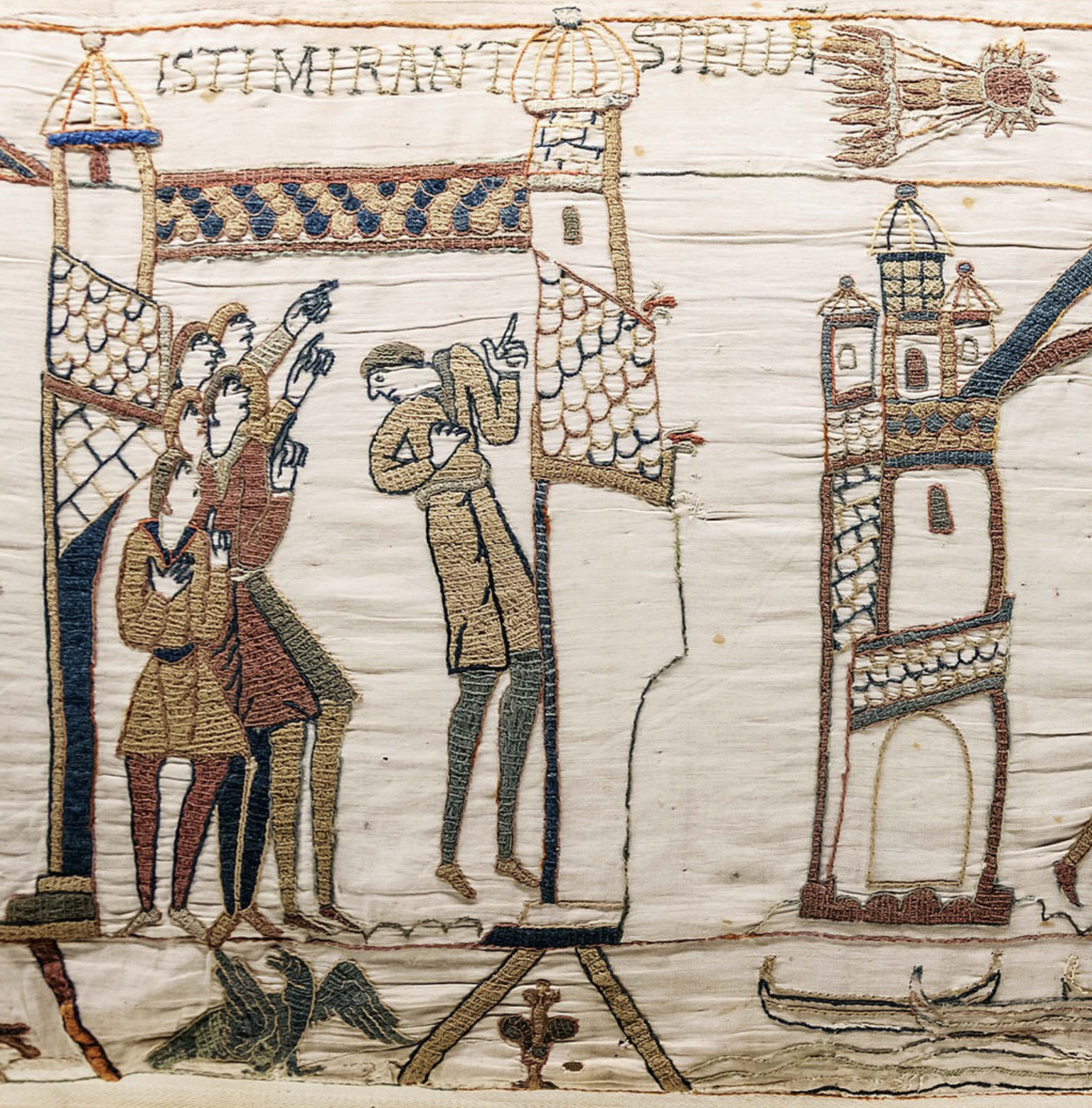}
\end{center}
\caption{Fragment from Bayeux Tapestry showing comet's Halley visit in 1066.}
\label{Bayeaux}
\end{figure}

In addition to the sources of the interstellar objects, it is important to study the sink process that limit their lifespan in the Galaxy, based on their size and galactic location, because these processes play a role in determining their distribution in phase space and size. The comparison between their predicted and observed distributions  can advance our understanding of the origin(s) of this new component of the interstellar medium and the clues it can unveil regarding planet formation. Sink process include size-dependent "dynamical heating" that could result in their ejection from the Galaxy (as their velocities increase with time due to perturbations by passing stars, molecular clouds, the Milky Way spiral arms, and star clusters); impacts with interstellar particles; disruption due to spin-up caused by the YORP effect; and evaporation. This is of interest regardless of the nature of the objects, but is particularly important if they are highly porous or contain a high fraction of highly volatile species, as it has been proposed for 1I/'Oumuamua (Seligman \& Laughlin \citeyear{2020ApJ...896L...8S}; Jackson \& Desch \citeyear{2021JGRE..12606706J}).  And generally, for all interstellar objects originating from planetary systems, it is important to understand how their ejection and entry (with a wide range of dynamical histories) can alter their physical properties (Raymond et al. \citeyear{2020ApJ...904L...4R}). 

Other observations that have been proposed to constrain the distribution in phase space and size and the nature of the interstellar objects are 
interstellar meteors, micro-meteorites and meteorites that, if detected,  would be able to probe the small end of their size distribution. However, Moro-Mart{\'{\i}}n (\citeyear{2018ApJ...866..131M}) compared the observed flux of meteorites and micrometeorites on Earth to those expected assuming the background density of interstellar objects discussed in Section \ref{numberdensity} and found that in all cases the observed fluxes are many orders of magnitude larger than expected. It is therefore unlikely that an interstellar meteorite is already part of the collected meteorite samples. Siraj \& Loeb (\citeyear{2019arXiv190603270S}, \citeyear {2021NewA...8401545S}) have suggested the detection of the first interstellar meteor corresponding to an object of $\sim$ 0.45 m in size, with an asymptotic speed of 42 \kms, 60 km/away from the velocity of the {\it Local Standard of Rest}. These type of objects, if identified pre-impact, could allow the study of their composition based on spectroscopy of their gaseous debris as they burn up in the atmosphere. They proposed using a network of all-sky cameras to determine the orbits and take spectra of a few hundred mm-sized meteors originating from the stellar halo in the Milky Way, that would be easy to distinguish from all other meteors because their characteristic asymptotic speed would be an order of magnitude higher. These observations could shed light on the characteristics of planetary system formation for the oldest stars that populate the Galactic halo.

It has also been proposed to observe an interstellar object from close range using a fly-by mission. The high asymptotic velocity expected for interstellar objects presents a challenge because of the increased $\Delta$v capability that the spacecraft would have to acquire during a relatively short period, but there is the argument that this could be addressed with scaled versions of existing technologies and a careful selection of the target (Seligman \& Laughlin \citeyear{2018AJ....155..217S}; Hibberd et al. \citeyear{2020AcAau.170..136H}, \citeyear{2021AcAau.189..584H}). The unpredictable frequency of interstellar objects would likely require to keep the spacecraft on a storage orbit until a suitable target is discovered, as is the plan for {\it Comet Interceptor}, a mission planned by the  {\it European Space Agency} for 2028, that will be parked at the Sun-Earth Lagrange point L2, but that might not have the $\Delta$v capability to reach an interstellar object. 

Figure \ref{Bayeaux} shows a fragment of the Balleaux Tapestry from 1066 showing bewildered observers of comet Halley. It summarizes well how much we have yet to learn about interstellar interlopers. But rather than evoking fear,  like it was the case back then, these interstellar objects evoke tremendous excitement and anticipation because they have truly opened a new era in astronomy in which we can dream of closely studying -- perhaps even holding in our hands and examining in our labs -- a fragment from another world beyond our solar system.


\begin{thebibliography}{99}
\bibitem[Adams(2010)]{2010ARA&A..48...47A} Adams, F.~C.\ 2010, \araa, 48, 47
\bibitem[Andrews \& Williams(2007)]{2007ApJ...671.1800A} Andrews, S.~M., \& Williams, J.~P.\ 2007, ApJ, 671, 1800 
\bibitem[Armitage(2010)]{2010apf..book.....A} Armitage, P.~J.\ 2010, Astrophysics of Planet Formation, by Philip J. Armitage, 294 pp. ISBN 978-0-521-88745-8 (hardback). Cambridge, UK: Cambridge University Press, 2010.
\bibitem[Bailer-Jones et al.(2018)]{2018AJ....156..205B} Bailer-Jones, C.~A.~L., Farnocchia, D., Meech, K.~J., et al.\ 2018, \aj, 156, 205. doi:10.3847/1538-3881/aae3eb
\bibitem[Bannister et al.(2017)]{2017ApJ...851L..38B} Bannister, M.~T., Schwamb, M.~E., Fraser, W.~C., et al.\ 2017, ApJL, 851, L38 
\bibitem[Bate \& Lor{\'e}n-Aguilar(2017)]{2017MNRAS.465.1089B} Bate, M.~R., \& Lor{\'e}n-Aguilar, P.\ 2017, \mnras, 465, 1089
\bibitem[Beichman et al.(2005)]{2005ApJ...626.1061B} Beichman, C.~A., Bryden, G., Gautier, T.~N., et al.\ 2005, \apj, 626, 1061. doi:10.1086/430059
\bibitem[Belbruno et al.(2012)]{2012AsBio..12..754B} Belbruno, E., Moro-Mart{\'\i}n, A., Malhotra, R., et al.\ 2012, Astrobiology, 12, 754
\bibitem[Belton et al.(2018)]{2018ApJ...856L..21B} Belton, M.~J.~S., Hainaut, O.~R., Meech, K.~J., et al.\ 2018, \apjl, 856, L21 
\bibitem[Bialy \& Loeb(2018)]{2018ApJ...868L...1B} Bialy, S. \& Loeb, A.\ 2018, \apjl, 868, L1. doi:10.3847/2041-8213/aaeda8
\bibitem[Birnstiel et al.(2011)]{2011A&A...525A..11B} Birnstiel, T., Ormel, C.~W., \& Dullemond, C.~P.\ 2011, \aap, 525, A11. doi:10.1051/0004-6361/201015228
\bibitem[Birnstiel et al.(2012)]{2012A&A...539A.148B} Birnstiel, T., Klahr, H., \& Ercolano, B.\ 2012, \aap, 539, A148. doi:10.1051/0004-6361/201118136
\bibitem[Blum \& Schr{\"a}pler(2004)]{2004PhRvL..93k5503B} Blum, J., \& Schr{\"a}pler, R.\ 2004, Physical Review Letters, 93, 115503 
\bibitem[Booth et al.(2018)]{2018MNRAS.475..167B} Booth, R.~A., Meru, F., Lee, M.~H., et al.\ 2018, \mnras, 475, 167
\bibitem[Bodewits et al.(2020)]{2020NatAs...4..867B} Bodewits, D., Noonan, J.~W., Feldman, P.~D., et al.\ 2020, Nature Astronomy, 4, 867. doi:10.1038/s41550-020-1095-2
\bibitem[Bolin et al.(2018)]{2018ApJ...852L...2B} Bolin, B.~T., Weaver, H.~A., Fernandez, Y.~R., et al.\ 2018, ApJL, 852, L2 
Borisov, G. 2019, MPEC, 2019-R106 (September 11)
\bibitem[Bottke et al.(2005)]{2005Icar..175..111B} Bottke, W.~F., Durda, D.~D., Nesvorn{\'y}, D., et al.\ 2005, \icarus, 175, 111. doi:10.1016/j.icarus.2004.10.026
\bibitem[Bottke et al.(2005)]{2005Icar..179...63B} Bottke, W.~F., Durda, D.~D., Nesvorn{\'y}, D., et al.\ 2005, \icarus, 179, 63. doi:10.1016/j.icarus.2005.05.017
\bibitem[Boyajian et al.(2016)]{2016MNRAS.457.3988B} Boyajian, T.~S., LaCourse, D.~M., Rappaport, S.~A., et al.\ 2016, MNRAS, 457, 3988 
\bibitem[Brasser et al.(2006)]{2006Icar..184...59B} Brasser, R., Duncan, M.~J., \& Levison, H.~F.\ 2006, Icar, 184, 59 
\bibitem[Brasser et al.(2010)]{2010A&A...516A..72B} Brasser, R., Higuchi, A., \& Kaib, N.\ 2010, A\&A, 516, A72 
\bibitem[Brasser et al.(2012)]{2012Icar..217....1B} Brasser, R., Duncan, M.~J., Levison, H.~F., Schwamb, M.~E., \& Brown, M.~E.\ 2012, Icar, 217, 1 
\bibitem[Brasser \& Morbidelli(2013)]{2013Icar..225...40B} Brasser, R., \& Morbidelli, A.\ 2013, Icar, 225, 40 
\bibitem[Burns et al.(1979)]{1979Icar...40....1B} Burns, J.~A., Lamy, P.~L., \& Soter, S.\ 1979, \icarus, 40, 1. doi:10.1016/0019-1035(79)90050-2
\bibitem[Carpenter et al.(2009)]{2009ApJS..181..197C} Carpenter, J.~M., Bouwman, J., Mamajek, E.~E., et al.\ 2009, \apjs, 181, 197. doi:10.1088/0067-0049/181/1/197
\bibitem[Cordiner et al.(2020)]{2020NatAs...4..861C} Cordiner, M.~A., Milam, S.~N., Biver, N., et al.\ 2020, Nature Astronomy, 4, 861. doi:10.1038/s41550-020-1087-2
\bibitem[Cumming et al.(2008)]{2008PASP..120..531C} Cumming, A., Butler, R. P.,  Marcy, G. W., Vogt, S. S., Wright, J. T. \ 2008, PASP, 120, 531
\bibitem[{\'C}uk(2018)]{2018ApJ...852L..15C} {\'C}uk, M.\ 2018, ApJL, 852, L15 
\bibitem[Do et al.(2018)]{2018ApJ...855L..10D} Do, A., Tucker, M.~A., \& Tonry, J.\ 2018, ApJL, 855, L10 
\bibitem[Dohnanyi(1969)]{1969JGR....74.2531D} Dohnanyi, J.~S.\ 1969, \jgr, 74, 2531. doi:10.1029/JB074i010p02531
\bibitem[Drahus et al.(2018)]{2018NatAs...2..407D} Drahus, M., Guzik, P., Waniak, W., et al.\ 2018, NatAs, 2, 407 
\bibitem[Dybczy{\'n}ski \& Kr{\'o}likowska(2018)]{2018A&A...610L..11D} Dybczy{\'n}ski, P.~A., \& Kr{\'o}likowska, M.\ 2018, A\&A, 610, L11 
\bibitem[Engelhardt et al.(2017)]{2017AJ....153..133E} Engelhardt, T., Jedicke, R., Vere{\v{s}}, P., et al.\ 2017, \aj, 153, 133. doi:10.3847/1538-3881/aa5c8a
\bibitem[Feng \& Jones(2018)]{2018ApJ...852L..27F} Feng, F., \& Jones, H.~R.~A.\ 2018, ApJL, 852, L27 
\bibitem[Fitzsimmons et al.(2018)]{2018NatAs...2..133F} Fitzsimmons, A., Snodgrass, C., Rozitis, B., et al.\ 2018, NatAs, 2, 133 
\bibitem[Flekk{\o}y et al.(2019)]{2019ApJ...885L..41F} Flekk{\o}y, E.~G., Luu, J., \& Toussaint, R.\ 2019, \apjl, 885, L41. doi:10.3847/2041-8213/ab4f78
\bibitem[Fraser et al.(2018)]{2018NatAs...2..383F} Fraser, W.~C., Pravec, P., Fitzsimmons, A., et al.\ 2018, NatAs, 2, 383 
\bibitem[Fulle et al.(2015)]{2015ApJ...802L..12F} Fulle, M., Della Corte, V., Rotundi, A., et al.\ 2015, \apjl, 802, L12 
\bibitem[Gaidos et al.(2017)]{2017RNAAS...1a..13G} Gaidos, E., Williams, J., \& Kraus, A.\ 2017, RNAAS, 1, 13 
\bibitem[Gaidos(2018)]{2018MNRAS.477.5692G} Gaidos, E.\ 2018, MNRAS, 477, 5692 
\bibitem[Gomes et al.(2005)]{2005Natur.435..466G} Gomes, R., Levison, H.~F., Tsiganis, K., et al.\ 2005, \nat, 435, 466. doi:10.1038/nature03676
\bibitem[Grishin et al.(2019)]{2019MNRAS.487.3324G} Grishin, E., Perets, H.~B., \& Avni, Y.\ 2019, \mnras, 487, 3324
\bibitem[Hansen \& Zuckerman(2017)]{2017RNAAS...1a..55H} Hansen, B., \& Zuckerman, B.\ 2017, RNAAS, 1, 55 
\bibitem[Hanse et al.(2018)]{2018MNRAS.473.5432H} Hanse, J., J{\'\i}lkov{\'a}, L., Portegies Zwart, S.~F., et al.\ 2018, \mnras, 473, 5432. doi:10.1093/mnras/stx2721
\bibitem[Harrison et al.(2005)]{2005Sci...310.1947H} Harrison, T.~M., Blichert-Toft, J., M{\"u}ller, W., et al.\ 2005, Science, 310, 1947. doi:10.1126/science.1117926
\bibitem[Hartmann (2008)]{Hartmann2008} Hartmann, L.\ 2008, Accretion Processes in Star Formation, (2nd ed., Cambridge Astrophysics). Cambridge: Cambridge University Press. doi:10.1017/CBO9780511552090
\bibitem[Hayashi(1981)]{1981PThPS..70...35H} Hayashi, C.\ 1981, Progress of Theoretical Physics Supplement, 70, 35
\bibitem[Hibberd et al.(2020)]{2020AcAau.170..136H} Hibberd, A., Hein, A.~M., \& Eubanks, T.~M.\ 2020, Acta Astronautica, 170, 136. doi:10.1016/j.actaastro.2020.01.018
\bibitem[Hibberd et al.(2021)]{2021AcAau.189..584H} Hibberd, A., Perakis, N., \& Hein, A.~M.\ 2021, Acta Astronautica, 189, 584. doi:10.1016/j.actaastro.2021.09.006
\bibitem[Hillenbrand et al.(2008)]{2008ApJ...677..630H} Hillenbrand, L.~A., Carpenter, J.~M., Kim, J.~S., et al.\ 2008, \apj, 677, 630. doi:10.1086/529027
\bibitem[Ivezi{\'c} et al.(2019)]{2019ApJ...873..111I} Ivezi{\'c}, {\v{Z}}., Kahn, S.~M., Tyson, J.~A., et al.\ 2019, \apj, 873, 111. doi:10.3847/1538-4357/ab042c
\bibitem[Jackson et al.(2018)]{2018MNRAS.477L..85J} Jackson, A.~P., Tamayo, D., Hammond, N., Ali-Dib, M., \& Rein, H.\ 2018, MNRAS, 477, L85 
\bibitem[Jackson \& Desch(2021)]{2021JGRE..12606706J} Jackson, A.~P. \& Desch, S.~J.\ 2021, Journal of Geophysical Research (Planets), 126, e06706. doi:10.1029/2020JE006706
\bibitem[Jewitt(2003)]{2003EM&P...92..465J} Jewitt, D.\ 2003, Earth Moon and Planets, 92, 465. doi:10.1023/B:MOON.0000031961.88202.60
\bibitem[Jewitt et al.(2017)]{2017ApJ...850L..36J} Jewitt, D., Luu, J., Rajagopal, J., et al.\ 2017, ApJL, 850, L36 
\bibitem[Jewitt \& Luu(2019)]{2019ApJ...886L..29J} Jewitt, D. \& Luu, J.\ 2019, \apjl, 886, L29. doi:10.3847/2041-8213/ab530b
\bibitem[Jewitt et al.(2020)]{2020ApJ...888L..23J} Jewitt, D., Hui, M.-T., Kim, Y., et al.\ 2020, \apjl, 888, L23
Jewitt, D. \& Moro-Martin, A., 2020, Scientific American (10/2020 issue). 
\bibitem[Jewitt et al.(2020)]{2020ApJ...896L..39J} Jewitt, D., Kim, Y., Mutchler, M., et al.\ 2020, \apjl, 896, L39. doi:10.3847/2041-8213/ab99cb
\bibitem[Johansen et al.(2004)]{2004A&A...417..361J} Johansen, A., Andersen, A.~C., \& Brandenburg, A.\ 2004, \aap, 417, 361
\bibitem[Johansen et al.(2007)]{2007Natur.448.1022J} Johansen, A., Oishi, J.~S., Mac Low, M.-M., et al.\ 2007, \nat, 448, 1022
\bibitem[Johansen \& Lambrechts(2017)]{2017AREPS..45..359J} Johansen, A., \& Lambrechts, M.\ 2017, AREPS, 45, 359 
\bibitem[Johnson et al.(2007a)]{2007ApJ...665..785J} Johnson, J.~A., Fischer, D.~A., Marcy, G.~W., et al.\ 2007, ApJ, 665, 785 
\bibitem[Kataoka et al.(2013)]{2013A&A...557L...4K} Kataoka, A., Tanaka, H., Okuzumi, S., et al.\ 2013, \aap, 557, L4. doi:10.1051/0004-6361/201322151
\bibitem[Kataoka(2017)]{2017ASSL..445..143K} Kataoka, A. 2017, Formation, Evolution and Dynamics of Yount Sytems, M. Pessah and O. Gressel (eds), Astrophysics and Space Science Library, Springer, Cham, pp. 143--159
\bibitem[Katyal et al.(2014)]{2014JQSRT.146..290K} Katyal, N., Banerjee, V., \& Puri, S.\ 2014, \jqsrt, 146, 290 
\bibitem[Katyal et al.(2014)]{2014JQSRT.146..290K} Katyal, N., Banerjee, V., \& Puri, S.\ 2014, \jqsrt, 146, 290 
\bibitem[Kennedy et al.(2018)]{2018MNRAS.476.4584K} Kennedy, G.~M., Bryden, G., Ardila, D., et al.\ 2018, \mnras, 476, 4584. doi:10.1093/mnras/sty492
\bibitem[Kenyon \& Bromley(2004)]{2004AJ....127..513K} Kenyon, S.~J. \& Bromley, B.~C.\ 2004, \aj, 127, 513. doi:10.1086/379854
\bibitem[Kiefer et al.(2014)]{2014Natur.514..462K} Kiefer, F., Lecavelier des Etangs, A., Boissier, J., et al.\ 2014, Natur, 514, 462 
\bibitem[Kleine et al.(2005)]{2005Sci...310.1671K} Kleine, T., Palme, H., Mezger, K., et al.\ 2005, Science, 310, 1671. doi:10.1126/science.1118842
\bibitem[Knight et al.(2017)]{2017ApJ...851L..31K} Knight, M.~M., Protopapa, S., Kelley, M.~S.~P., et al.\ 2017, \apjl, 851, L31. doi:10.3847/2041-8213/aa9d81
\bibitem[Kroupa et al.(1993)]{1993MNRAS.262..545K} Kroupa, P., Tout, C.~A., \& Gilmore, G.\ 1993, MNRAS, 262, 545 
\bibitem[Levine et al.(2021)]{2021arXiv210811194L} Levine, W.~G., Cabot, S.~H.~C., Seligman, D., et al.\ 2021, arXiv:2108.11194
\bibitem[Levison et al.(2010)]{2010Sci...329..187L} Levison, H.~F., Duncan, M.~J., Brasser, R., et al.\ 2010, Science, 329, 187. doi:10.1126/science.1187535
\bibitem[Luu et al.(2020)]{2020ApJ...900L..22L} Luu, J.~X., Flekk{\o}y, E.~G., \& Toussaint, R.\ 2020, \apjl, 900, L22. doi:10.3847/2041-8213/abafa7
\bibitem[Mamajek(2017)]{2017RNAAS...1a..21M} Mamajek, E.\ 2017, RNAAS, 1, 21 
\bibitem[Marcy et al.(2005)]{2005PThPS.158...24M} Marcy, G., Butler, R.~P., Fischer, D., et al.\ 2005, Progress of Theoretical Physics Supplement, 158, 24 
\bibitem[McGlynn \& Chapman(1989)]{1989ApJ...346L.105M} McGlynn, T.~A. \& Chapman, R.~D.\ 1989, \apjl, 346, L105. doi:10.1086/185590
\bibitem[McNeill et al.(2018)]{2018ApJ...857L...1M} McNeill, A., Trilling, D.~E., \& Mommert, M.\ 2018, \apjl, 857, L1 
\bibitem[Meech et al.(2017)]{2017Natur.552..378M} Meech, K.~J., Weryk, R., Micheli, M., et al.\ 2017, Natur, 552, 378 
\bibitem[Meyer et al.(2008)]{2008ApJ...673L.181M} Meyer, M.~R., Carpenter, J.~M., Mamajek, E.~E., et al.\ 2008, \apjl, 673, L181. doi:10.1086/527470
\bibitem[Meng et al.(2014)]{2014Sci...345.1032M} Meng, H.~Y.~A., Su, K.~Y.~L., Rieke, G.~H., et al.\ 2014, Science, 345, 1032. doi:10.1126/science.1255153
\bibitem[Micheli et al.(2018)]{2018Natur.559..223M} Micheli, M., Farnocchia, D., Meech, K.~J., et al.\ 2018, Natur, 559, 223 
\bibitem[Mojzsis et al.(1996)]{1996Natur.384...55M} Mojzsis, S.~J., Arrhenius, G., McKeegan, K.~D., et al.\ 1996, \nat, 384, 55. doi:10.1038/384055a0
\bibitem[Morbidelli et al.(2000)]{2000M&PS...35.1309M} Morbidelli, A., Chambers, J., Lunine, J.~I., et al.\ 2000, \maps, 35, 1309. doi:10.1111/j.1945-5100.2000.tb01518.x
\bibitem[Morbidelli et al.(2012)]{2012AREPS..40..251M} Morbidelli, A., Lunine, J.~I., O'Brien, D.~P., et al.\ 2012, Annual Review of Earth and Planetary Sciences, 40, 251. doi:10.1146/annurev-earth-042711-105319
\bibitem[Moro-Mart{\'{\i}}n et al.(2009)]{2009ApJ...704..733M} Moro-Mart{\'{\i}}n, A., Turner, E.~L., \& Loeb, A.\ 2009, ApJ, 704, 733 
\bibitem[Moro-Mart{\'{\i}}n(2013)]{2013pss3.book..431M} Moro-Mart{\'{\i}}n, A.\ 2013, Planets, Stars and Stellar Systems.~Volume 3: Solar and Stellar Planetary Systems, T. D. Oswalt, L. M. French, P. Kalas (eds.), Springer Science+Business Media, Dordrecht, pp. 431-487
\bibitem[Moro-Mart{\'\i}n et al.(2015)]{2015ApJ...801..143M} Moro-Mart{\'\i}n, A., Marshall, J.~P., Kennedy, G., et al.\ 2015, \apj, 801, 143. doi:10.1088/0004-637X/801/2/143
\bibitem[Moro-Mart{\'{\i}}n(2018)]{2018ApJ...866..131M} Moro-Mart{\'{\i}}n, A.\ 2018, \apj, 866, 131
\bibitem[Moro-Mart{\'\i}n(2019a)]{2019AJ....157...86M} Moro-Mart{\'\i}n, A.\ 2019, \aj, 157, 86
\bibitem[Moro-Mart{\'\i}n(2019b)]{2019ApJ...872L..32M} Moro-Mart{\'\i}n, A.\ 2019, \apjl, 872, L32
\bibitem[Moro-Mart{\'\i}n \& Norman(2021)]{2021arXiv211015366M} Moro-Mart{\'\i}n, A. \& Norman, C.\ 2021, arXiv:2110.15366
\bibitem[Nesvorn{\'y}(2018)]{2018ARA&A..56..137N} Nesvorn{\'y}, D.\ 2018, \araa, 56, 137. doi:10.1146/annurev-astro-081817-052028
\bibitem[Okuzumi et al.(2009a)]{2009ApJ...707.1247O} Okuzumi, S., Tanaka, H., \& Sakagami, M.-. aki .\ 2009, \apj, 707, 1247
\bibitem[Okuzumi(2009b)]{2009ApJ...698.1122O} Okuzumi, S.\ 2009, \apj, 698, 1122
\bibitem[Okuzumi et al.(2012)]{2012ApJ...752..106O} Okuzumi, S., Tanaka, H., Kobayashi, H., et al.\ 2012, \apj, 752, 106. doi:10.1088/0004-637X/752/2/106
\bibitem['Oumuamua ISSI Team et al.(2019)]{2019NatAs...3..594O} 'Oumuamua ISSI Team, Bannister, M.~T., Bhandare, A., et al.\ 2019, Nature Astronomy, 3, 594. doi:10.1038/s41550-019-0816-x
\bibitem[Pfalzner, \& Bannister(2019)]{2019ApJ...874L..34P} Pfalzner, S., \& Bannister, M.~T.\ 2019, \apjl, 874, L34
\bibitem[Pfalzner et al.(2021)]{2021arXiv210608580P} Pfalzner, S., Paterson, D., Bannister, M.~T., et al.\ 2021, arXiv:2106.08580
\bibitem[Portegies Zwart et al.(2018)]{2018MNRAS.479L..17P} Portegies Zwart, S., Torres, S., Pelupessy, I., B{\'e}dorf, J., \& Cai, M.~X.\ 2018, MNRAS, 479, L17 
\bibitem[Rafikov(2018a)]{2018ApJ...861...35R} Rafikov, R.~R.\ 2018, ApJ, 861, 35 
\bibitem[Rafikov(2018b)]{2018ApJ...867L..17R} Rafikov, R.~R.\ 2018b, \apjl, 867, L17 
\bibitem[Raymond et al.(2009)]{2009Icar..203..644R} Raymond, S.~N., O'Brien, D.~P., Morbidelli, A., et al.\ 2009, \icarus, 203, 644. doi:10.1016/j.icarus.2009.05.016
\bibitem[Raymond et al.(2011)]{2011A&A...530A..62R} Raymond, S.~N., Armitage, P.~J., Moro-Mart{\'\i}n, A., et al.\ 2011, \aap, 530, A62. doi:10.1051/0004-6361/201116456
\bibitem[Raymond et al.(2012)]{2012A&A...541A..11R} Raymond, S.~N., Armitage, P.~J., Moro-Mart{\'\i}n, A., et al.\ 2012, \aap, 541, A11. doi:10.1051/0004-6361/201117049
\bibitem[Raymond et al.(2018b)]{2018MNRAS.476.3031R} Raymond, S.~N., Armitage, P.~J., Veras, D., Quintana, E.~V., \& Barclay, T.\ 2018, MNRAS, 476, 3031 
\bibitem[Raymond et al.(2018a)]{2018ApJ...856L...7R} Raymond, S.~N., Armitage, P.~J., \& Veras, D.\ 2018, ApJL, 856, L7 
\bibitem[Raymond et al.(2020)]{2020ApJ...904L...4R} Raymond, S.~N., Kaib, N.~A., Armitage, P.~J., et al.\ 2020, \apjl, 904, L4. doi:10.3847/2041-8213/abc55f
\bibitem[Schlichting et al.(2013)]{2013AJ....146...36S} Schlichting, H.~E., Fuentes, C.~I., \& Trilling, D.~E.\ 2013, \aj, 146, 36. doi:10.1088/0004-6256/146/2/36
\bibitem[Schopf(1993)]{1993Sci...260..640S} Schopf, J.~W.\ 1993, Science, 260, 640. doi:10.1126/science.260.5108.640
\bibitem[Sekanina(2019)]{2019arXiv190108704S} Sekanina, Z.\ 2019, arXiv:1901.08704
\bibitem[Seligman \& Laughlin(2018)]{2018AJ....155..217S} Seligman, D. \& Laughlin, G.\ 2018, \aj, 155, 217. doi:10.3847/1538-3881/aabd37
\bibitem[Seligman et al.(2019)]{2019ApJ...876L..26S} Seligman, D., Laughlin, G., \& Batygin, K.\ 2019, \apjl, 876, L26. doi:10.3847/2041-8213/ab0bb5
\bibitem[Seligman \& Laughlin(2020)]{2020ApJ...896L...8S} Seligman, D. \& Laughlin, G.\ 2020, \apjl, 896, L8. doi:10.3847/2041-8213/ab963f
\bibitem[Siraj \& Loeb(2019)]{2019arXiv190603270S} Siraj, A. \& Loeb, A.\ 2019, arXiv:1906.03270
\bibitem[Siraj \& Loeb(2021)]{2021NewA...8401545S} Siraj, A. \& Loeb, A.\ 2021, \na, 84, 101545. doi:10.1016/j.newast.2020.101545
\bibitem[Siraj \& Loeb(2021)]{2021arXiv210314032S} Siraj, A. \& Loeb, A.\ 2021, arXiv:2103.14032
\bibitem[Shu et al.(1987)]{1987ARA&A..25...23S} Shu, F.~H., Adams, F.~C., \& Lizano, S.\ 1987, \araa, 25, 23. doi:10.1146/annurev.aa.25.090187.000323
\bibitem[Su et al.(2006)]{2006ApJ...653..675S} Su, K.~Y.~L., Rieke, G.~H., Stansberry, J.~A., et al.\ 2006, \apj, 653, 675. doi:10.1086/508649
\bibitem[Su et al.(2019)]{2019AJ....157..202S} Su, K.~Y.~L., Jackson, A.~P., G{\'a}sp{\'a}r, A., et al.\ 2019, \aj, 157, 202. doi:10.3847/1538-3881/ab1260
\bibitem[Suyama et al.(2008)]{2008ApJ...684.1310S} Suyama, T., Wada, K., \& Tanaka, H.\ 2008, \apj, 684, 1310 
\bibitem[Trilling et al.(2018)]{2018AJ....156..261T} Trilling, D.~E., Mommert, M., Hora, J.~L., et al.\ 2018, \aj, 156, 261 
\bibitem[Veras et al.(2011)]{2011MNRAS.417.2104V} Veras, D., Wyatt, M.~C., Mustill, A.~J., Bonsor, A., \& Eldridge, J.~J.\ 2011, MNRAS, 417, 2104 
\bibitem[Veras \& Tout(2012)]{2012MNRAS.422.1648V} Veras, D., \& Tout, C.~A.\ 2012, MNRAS, 422, 1648 
\bibitem[Veras et al.(2014)]{2014MNRAS.437.1127V} Veras, D., Evans, N.~W., Wyatt, M.~C., et al.\ 2014, \mnras, 437, 1127. doi:10.1093/mnras/stt1905
\bibitem[Weidenschilling(1977)]{1977MNRAS.180...57W} Weidenschilling, S.~J.\ 1977, \mnras, 180, 57
\bibitem[Weidenschilling(1980)]{1980Icar...44..172W} Weidenschilling, S.~J.\ 1980, \icarus, 44, 172
\bibitem[Welsh \& Montgomery(2015)]{2015AdAst2015E..26W} Welsh, B.~Y., \& Montgomery, S.~L.\ 2015, Advances in Astronomy, 2015, 980323 
\bibitem[Williams(2017)]{Williams2017}Williams, G., Minor Planet Electronic Circular 2017-U181 (October 25)
\bibitem[Windmark et al.(2012)]{2012A&A...544L..16W} Windmark, F., Birnstiel, T., Ormel, C.~W., et al.\ 2012, \aap, 544, L16. doi:10.1051/0004-6361/201220004
\bibitem[Winn \& Fabrycky(2015)]{2015ARA&A..53..409W} Winn, J.~N., \& Fabrycky, D.~C.\ 2015, ARA\&A, 53, 409 
\bibitem[Wyatt(2008)]{2008ARA&A..46..339W} Wyatt, M.~C.\ 2008, \araa, 46, 339. doi:10.1146/annurev.astro.45.051806.110525
\bibitem[Wyatt et al.(2017)]{2017MNRAS.464.3385W} Wyatt, M.~C., Bonsor, A., Jackson, A.~P., Marino, S., \& Shannon, A.\ 2017, MNRAS, 464, 3385 
\bibitem[Xie et al.(2010)]{2010ApJ...724.1153X} Xie, J.-W., Payne, M.~J., Th{\'e}bault, P., et al.\ 2010, \apj, 724, 1153
\bibitem[Ye et al.(2017)]{2017ApJ...851L...5Y} Ye, Q.-Z., Zhang, Q., Kelley, M.~S.~P., et al.\ 2017, \apjl, 851, L5. doi:10.3847/2041-8213/aa9a34
\bibitem[Youdin \& Shu(2002)]{2002ApJ...580..494Y} Youdin, A.~N. \& Shu, F.~H.\ 2002, \apj, 580, 494. doi:10.1086/343109
\bibitem[Youdin \& Goodman(2005)]{2005ApJ...620..459Y} Youdin, A.~N. \& Goodman, J.\ 2005, \apj, 620, 459. doi:10.1086/426895
\bibitem[Zhang(2018)]{2018ApJ...852L..13Z} Zhang, Q.\ 2018, ApJL, 852, L13 
\bibitem[Zhang \& Lin(2020)]{2020NatAs...4..852Z} Zhang, Y. \& Lin, D.~N.~C.\ 2020, Nature Astronomy, 4, 852. doi:10.1038/s41550-020-1065-8
\bibitem[Zsom et al.(2010)]{2010A&A...513A..57Z} Zsom, A., Ormel, C.~W., G{\"u}ttler, C., et al.\ 2010, \aap, 513, A57. doi:10.1051/0004-6361/200912976

\end{thebibliography}
\end{document}